\documentclass[a4 paper, 11pt]{article}
\pdfoutput=1
\usepackage{comment}
\usepackage{jheppub}
\usepackage{subcaption}
\usepackage{amsmath}
\usepackage{mathrsfs}
\usepackage{amssymb}
\usepackage{amscd}
\usepackage{dsfont}
\usepackage{enumerate}
\usepackage{amsfonts}
\usepackage{epsfig}
\usepackage{mathtools}
\usepackage{yfonts}
\usepackage{bbold}
\usepackage{breqn}
\usepackage[utf8]{inputenc}
\usepackage[english]{babel}
\usepackage{graphicx}
\usepackage{empheq}
\usepackage{relsize}
\usepackage{braket}

\usepackage{graphicx}
\usepackage{tikz}
\usetikzlibrary{arrows.meta,decorations.markings}

\usepackage{cancel}

\usepackage{jheppub}
\usepackage{mathtools}
\usepackage{float}
\usepackage{tikz-cd}

\date{}

\newcommand{\cd}{\cdots}

\newcommand{\bp}{\bar{\partial}}

 \usepackage[normalem]{ulem}
\usepackage[textsize=scriptsize,textwidth=2.5cm]{todonotes}

\let\exporig\exp
\usepackage{physics}
\usepackage{enumitem}
\let\exp\exporig

\affiliation[\ensuremath{\gamma}]{Yau Mathematical Sciences Center, Tsinghua University, Beijing 100084, China}
\affiliation[\ensuremath{\tau}]{Department of Mathematical Sciences, Tsinghua University, Beijing 100084, China}

\usepackage{xspace}

\allowdisplaybreaks
\newcommand{\z}{\left}
\newcommand{\y}{\right}
\newcommand{\p}{\partial}

\newcommand{\cy}{\mathrm{cy}}
\newcommand{\pl}{\mathrm{pl}}

\newcommand{\ren}{\mathrm{ren}}
\newcommand{\bare}{\mathrm{bare}}

\newcommand{\loc}{\mathrm{loc}}

\newcommand{\JT}{J\bar T}
\newcommand{\TT}{T\bar T}
\newcommand{\Dr}{\operatorname{Dr}}

\newcommand{\KE}{\operatorname{KE}}

\title{
$J\bar J$-deformation as a Riemann bilinear dressing
}

\author{Kangning Liu$^{\gamma,\tau}$}

\emailAdd{lkn22@mails.tsinghua.edu.cn}

\abstract
{ 

We propose a reformulation of the conformal perturbation theory of the correlation functions in $J\bar J$-deformed CFTs as a dressing on the deformed operators, that matches both bare and renormalized perturbation theory. The key is to use the Riemann bilinear identity to convert the deformation into a dressing and a large-cycle integral for higher genus. Based on the proposal, we calculate the deformation of partition functions on the torus and higher genus Riemann surfaces, which can be written as kernel integrals that preserve modular invariance or covariance. We also calculate the flow of the conformal weights and conserved charges along the deformation. Based on this flow and the modular $S$-transformation, we propose a criterion for constructing dressed operators. We test our formalism and results by studying the $O(2,2)$ theories and strings on the TsT background.

}

\begin{document}
\maketitle

\section{Introduction}

Exactly marginal deformations provide one of the most useful ways of moving on
the conformal manifold of two-dimensional conformal field theories. Among them,
current-current deformations are particularly tractable \cite{Borsato:2023dis}. Given a pair of chiral
and anti-chiral currents, one can perturb the action by an operator of the form
\begin{equation}
    \frac{\p S_\lambda}{\p\lambda}
    =
    \int d^2z~J_1^{(\lambda)}(z)\bar J_2^{(\lambda)}(\bar z).
\end{equation}
Such deformations are usually called $J\bar J$-deformations. They are special
because the perturbing operator has dimension $(1,1)$ and, for abelian current
algebras, it can remain exactly marginal at a finite deformation parameter. The basic criterion was formulated in \cite{Chaudhuri:1988qb}: a bilinear of chiral and anti-chiral currents gives an
integrably marginal deformation when the currents involved generate mutually
commuting current algebras. Later analyses, including the study of current-current perturbations in CFTs and WZW models, further clarified that the
chiral currents and the stress tensor can be transported along the deformation,
so that the deformed theory continues to possess the corresponding holomorphic
and anti-holomorphic symmetries
\cite{Forste:2003km,Fredenhagen:2007rx,Borsato:2024cct}.

A familiar example is the radius deformation of a compact free boson. Changing
the radius is generated by a bilinear of left- and right-moving $U(1)$ currents,
and the effect of the deformation is to rotate the momentum-winding lattice.
More generally, current-current deformations of toroidal CFTs are naturally
organized by Narain compactifications and their $O(d,d)$ duality structure
\cite{Narain:1985jj,Narain:1986am,Giveon:1994fu}. From the target-space point
of view, finite abelian current-current deformations are closely related to
T-duality--shift--T-duality transformations, or TsT transformations. This
perspective makes it clear that the deformation is not merely a perturbative
operation on the Lagrangian: it can also be understood as changing the gluing
conditions, the charge lattice, and the spectrum of mutually local operators.
This point of view has been useful in WZW models, integrable sigma models, and
solvable string backgrounds
\cite{Hassan:1992gi,Lunin:2005jy,Sfondrini:2019smd}.

Current-current deformations have also become important in holography. The
irrelevant $\TT$ deformation of two-dimensional QFTs is solvable in a remarkable
sense \cite{Zamolodchikov:2004ce,Smirnov:2016lqw,Cavaglia:2016oda}, and its
holographic interpretation has led to a large body of work on finite-cutoff
AdS$_3$ gravity and related non-AdS geometries \cite{McGough:2016lol}. In
string theory on AdS$_3$ with NS-NS flux, the single-trace version of the
$\TT$ deformation is realized as an exactly marginal current-current deformation
of the worldsheet theory \cite{Giveon:2017nie,Giveon:2017myj}. The same
mechanism extends to $\JT$, $T\bar J$, and $J\bar J$-type deformations, and
their finite target-space realizations are often TsT transformations
\cite{Chakraborty:2019mdf,Apolo:2019zai,Hashimoto:2019wct,Apolo:2021wcn,Guica:2017lia, Guica:2021fkv}.
This relation has been sharpened by studies of spectra, partition functions,
correlation functions, and asymptotic symmetries in the TsT/$\TT$ correspondence
\cite{Cui:2023jrb,Du:2024bqk}. Thus a deformation that is irrelevant from the
boundary QFT point of view can be represented by a marginal and controllable
deformation on the worldsheet. This is one of the main reasons why
$J\bar J$-deformations are important in holography: they provide a bridge
between solvable worldsheet CFT deformations, TsT-transformed geometries, and
single-trace irrelevant deformations of the spacetime dual.

Despite these geometric and sigma-model interpretations, there remains a useful
CFT question: how should one describe the deformation directly at the level of
correlation functions and operators? In conformal perturbation theory, the
first-order variation of a correlation function contains an area integral of the
perturbing operator. This integral is singular near operator insertions and
requires a regularization and renormalization prescription. Moreover, the
operators in the deformed theory are not literally the same operators as those
in the undeformed theory; their conformal weights, charges, and mutual locality
conditions generally depend on $\lambda$. Therefore, to compare different
points on the conformal manifold, one needs an operator map between the
deformed and undeformed theories. Related operator-level and correlation-function
perspectives on current-current and solvable irrelevant deformations include
\cite{Moriwaki:2020zyl,Giribet:2020mkz,Cardy:2019qao,Kruthoff:2020hsi}.

The main idea of this work is to rewrite the conformal perturbation theory of
$J\bar J$-deformed CFTs as a dressing of operators. The key technical tool is
the Riemann bilinear identity. Since $J_1(z)dz$ and
$\bar J_2(\bar z)d\bar z$ are holomorphic and anti-holomorphic one-forms away
from insertions, the area integral of $J_1\bar J_2$ can be converted into cycle
integrals. On the sphere, these cycle integrals localize around the inserted
operators and produce a bare dressing. On higher-genus Riemann surfaces, there
is an additional large-cycle contribution, which is responsible for the flow of
the partition function. After renormalization, the dressing becomes a map
\begin{equation}
    \Dr_\lambda^{\ren}: \mathcal H_\lambda \longrightarrow \mathcal H_0
\end{equation}
from the operator space of the deformed theory to that of the undeformed theory.
A central property of this map is that it sends the deformed currents and stress
tensor back to the undeformed ones,
\begin{equation}
    \Dr_\lambda^{\ren}(J_1^{(\lambda)})=J_1^{(0)},
    \qquad
    \Dr_\lambda^{\ren}(T^{(\lambda)})=T^{(0)}.
\end{equation}
Thus the deformation can be viewed as changing the spectrum of dressed operators
inside a fixed undeformed CFT, while keeping the stress tensor used to evaluate
their correlation functions fixed.

This viewpoint gives a unified way to treat local operators, partition
functions, modular properties, and explicit string theory examples. We first
derive the dressing from the Riemann bilinear identity. We then compute the
deformation of torus and higher-genus partition functions, both without and
with current self-interactions. In the absence of self-interactions, the flow is
a heat equation in the chemical potentials. With self-interactions, the flow is
modified by first-order drift terms controlled by the levels $\kappa$ and
$\bar\kappa$. We show that the resulting kernels preserve modular invariance or
modular covariance after the appropriate non-holomorphic completion. We also
derive the flow of conformal weights and charges and use the modular
$S$-transformation to formulate a criterion for constructing dressed operators
as endpoint operators of topological defects.

Let us summarize the organization of the paper. In Section
\ref{secgeoRBI}, we review the geometric Riemann bilinear identity on a compact
Riemann surface, including the case of meromorphic differentials with poles. We
then apply this identity to the first-order $J\bar J$ deformation of CFT
correlation functions and explain how the area integral becomes a sum of local
dressing terms and, on higher-genus surfaces, large-cycle terms. We also discuss
the role of contact divergences and formulate the renormalized dressing map.
The subsection on a single free boson provides a simple test case: the dressing
reproduces the familiar radius deformation and clarifies how the spectrum of
mutually local vertex operators is transported along the deformation.

In Section \ref{sec:partitionfunc}, we study partition functions with chemical
potentials. We begin with the case in which the currents have no
self-interactions. On the torus, the Riemann bilinear identity reduces the
deformation to a heat equation in the chemical potentials, and the integrated
solution is a Gaussian kernel acting on the undeformed partition function. We
then generalize this result to higher genus, where the period matrix and its
imaginary part replace the torus modulus $\tau_2$. We show that the kernel
preserves modular invariance. We then include self-interactions, namely
non-zero current levels $\kappa$ and $\bar\kappa$. The Ward identity now
contains second-kind differentials, and the partition function obeys a
modified flow equation with drift terms. We derive the corresponding
higher-genus formula, discuss modular covariance and non-holomorphic
completions, and write the integrated answer as a harmonic-oscillator-type
kernel. Finally, we use the flow equation to extract the deformation of
individual conformal weights and charges. This leads to a simple
state-by-state flow, and the modular $S$-transformation suggests a criterion
for constructing dressed operators as endpoint operators of defects.

In Section \ref{app:rank-one-free-bosons}, we test the formalism in
$O(2,2)$ theories. We consider two compact bosons with a general rank-one
current-current deformation. We distinguish the non-flat coordinate appearing
linearly in the sigma-model action from the flat coordinate of conformal
perturbation theory. We show that the bare dressing of the elementary real
bosons is already the renormalized dressing in the $O(2,2)$ subsystem, and we
check explicitly that it maps the deformed OPE metric, currents, conformal
weights, and charges to the expected undeformed quantities. We then discuss the
partition function and verify the flow equation state by state. In the
no-self-interaction examples, we emphasize that the absence of current levels
does not necessarily imply a trivial flat coordinate or a trivial dressing; the
mixed OPE coefficients can still run and affect the operator map. We also
illustrate the additional subtleties in dressing non-exponential primary
operators and descendants.

In Section \ref{subsec:TsT-real-O22-corrected}, we apply the construction to
strings on a TsT-deformed AdS$_3$ background. We use the standard worldsheet
formulation of strings on AdS$_3$ and the $SL(2,\mathbb R)$ WZW model
\cite{Giveon:1998ns,Maldacena:2000hw,Maldacena:2000kv,Maldacena:2001km}, and
follow the Wakimoto conventions of \cite{Knighton:2023mhq}. We rewrite the
Wakimoto variables in a real $O(2,2)\oplus O(1,1)$ basis and identify the TsT
deformation as a null current-current deformation in the $O(2,2)$ sector. The
renormalized dressing of the real bosons then gives a direct operator-level
description of the deformed worldsheet theory. We show that the deformed stress
tensor is dressed back to the undeformed one and construct the corresponding
deformed vertex operators. Their conformal weights reproduce the expected charge
flow and agree with the known TsT/single-trace deformation picture.

\section{Perturbation theory as a Riemann bilinear identity}
We define $J\bar J$-deformation as the deformation of the correlation functions via perturbation theory. To be more precise, let us consider a CFT with action $S_0$ and $U(1)\times U(1)$ symmetry generated by $J_1(z)$ and $\bar J_2(\bar z)$. We consider the $J\bar J$-deformation defined by
\begin{equation}
\begin{aligned}
    \frac{\p}{\p \lambda} S_\lambda = \int d^2z ~ J_{1}^{(\lambda)}(z) \bar J^{(\lambda)}_{2}(\bar z) ,\qquad d^2z : = \frac{i}{2}dz\wedge d\bar z.
\end{aligned}
\end{equation}
The OPE of the currents in the undeformed theory is parametrized by fixed levels $\kappa,\bar \kappa$:
\begin{equation}
    J_{1}^{(\lambda)}(z) J_{1}^{(\lambda)}(w) \sim - \frac{\kappa}{(z-w)^2} ,\quad \bar J_{2}^{(\lambda)}(z) \bar J_{2}^{(\lambda)}(w) \sim - \frac{\bar \kappa}{(z-w)^2}.
\end{equation}
We will mainly focus on the no self-interaction case $\kappa=\bar \kappa=0$ and the case with self-interaction $\kappa,~ \bar \kappa \neq 0$. This deformation parameter $\lambda$ is a flat coordinate on the curve of $J_1\bar J_2$-deformation embedded in the moduli space of exactly marginal deformations. 

Since we only consider a $U(1)\times U(1)$ abelian current-current deformation, the deformation trivially satisfies the integrably-marginal condition proposed in \cite{Chaudhuri:1988qb}. Moreover, it is argued in \cite{Fredenhagen:2007rx} that the currents $J_1$, $\bar J_2$, and the stress tensor $T(z)$ will remain chiral symmetries of the theory. Thus, the deformation will be an exact marginal deformation even for finite $\lambda$. We will provide several arguments for these facts later. 

The deformation of correlation functions for a series of insertions $X$ can be defined perturbatively
\begin{equation}\label{deformnaive}
\begin{aligned}
  & \langle X^{(\lambda+\delta\lambda)} \rangle_{\lambda+\delta \lambda} = \langle X^{(\lambda+\delta\lambda)} \z(1-\delta \lambda \int d^2z ~ J_{1}^{(\lambda)}(z) \bar J^{(\lambda)}_{2}(\bar z) \y) \rangle_\lambda +  \mathcal{O}(\delta\lambda^2).
\end{aligned}
\end{equation}
The important subtlety is that $X^{(\lambda+\delta \lambda)} \neq X^{(\lambda)}$, since the spectrum of mutually local operators usually depends on $\lambda$. Moreover, the perturbation calculation of the RHS usually encounters UV divergences. We will assume that there exists a regularization and renormalization scheme in the conformal perturbation theory \cite{cardy1996scaling}, so that
\begin{equation}\label{eq1.4}
    \langle X^{(\lambda+\delta\lambda)}\z(1-\delta \lambda \int d^2z ~ J_{1}^{(\lambda)}(z) \bar J^{(\lambda)}_{2}(\bar z) \y) \rangle_{\lambda,\text{renormalized}} = \langle X^{(\lambda+\delta\lambda)} \rangle_{\lambda+\delta\lambda} + \mathcal{O}(\delta\lambda^2).
\end{equation}
On the other hand, before the renormalization, the deformation area integral can be rewritten as a cycle integral that locally\footnote{By local, we mean the cycle integral can be further localized by introducing primitives of currents. } links the insertions $X^{(\lambda+\delta \lambda)}$ using the Riemann bilinear identity. We will refer to this local cycle integral as a (bare) ``dressing" on the operators, denoted as $\Dr_{\delta\lambda}$. The renormalized dressing will be denoted as $\Dr^{\operatorname{ren}}_{\delta \lambda}$, then \eqref{eq1.4} suggests
\begin{equation}\label{eq2.5}
    \langle \Dr^{\operatorname{ren}}_{\delta \lambda}(X^{(\lambda+\delta \lambda)} ) \rangle_{\lambda} = \langle X^{(\lambda+\delta\lambda)} \rangle_{\lambda+\delta\lambda}.
\end{equation}
That is, we can use some operators defined in the $S_\lambda$-theory to calculate the correlation functions in the $S_{\lambda+\delta \lambda}$-theory. 

In this section, we will introduce the Riemann bilinear identity as a tool to study the $J\bar J$-deformations. Note that similar techniques have also been applied in \cite{Cardy:2019qao} and \cite{Callebaut:2019omt} for studying correlation functions in $T\bar T$-deformed theory and non-critical strings. For self-containedness, let us first review the proof of the geometric Riemann bilinear identity and then apply it to the $J\bar J$ deformed CFTs. 

\subsection{Riemann bilinear identity}\label{secgeoRBI}
Consider a Riemann surface $\Sigma$ of genus $g$, and introduce a basis $\{A_i,B_i\}_{i=1}^g$ of $H_1(\Sigma,\mathbb Z)$ so that $A_i \cap B_j = \delta_{ij}$, $A_i\cap A_j = B_i \cap B_j = 0$. For holomorphic 1-forms $\omega_1,\omega_2$, we have
\begin{equation}\label{holoRBI}
    \int_{\Sigma} \omega_1 \wedge \bar \omega_2 = \sum_{j=1}^g \z( \oint_{A_j} \omega_1 \oint_{B_j} \bar\omega_2  - \oint_{B_j} \omega_1 \oint_{A_j} \bar\omega_2 \y).
\end{equation}
This is called the Riemann bilinear identity (RBI). 

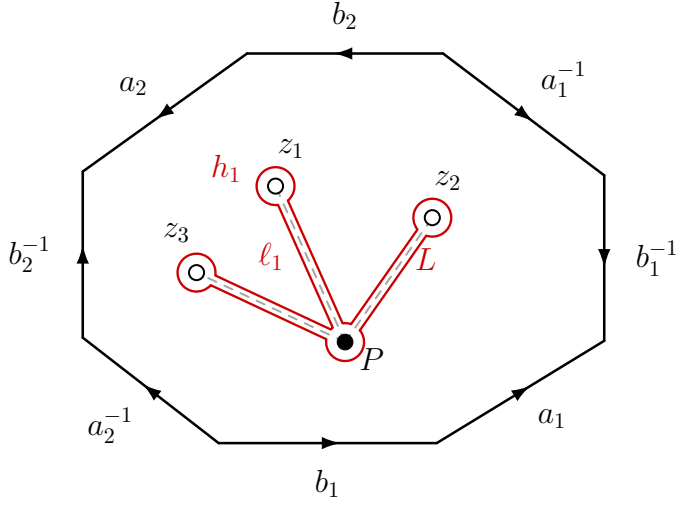
\begin{figure}[htbp]
\centering
\resizebox{0.6\linewidth}{!}{%
\begin{tikzpicture}[
  line cap=round,
  line join=round,
  every node/.style={font=\Large}
]
\tikzset{
  boundary/.style={
    black,
    line width=1.15pt,
    postaction={decorate},
    decoration={
      markings,
      mark=at position 0.56 with {\arrow{Latex[length=3.0mm,width=2.0mm]}}
    }
  },
  centerpath/.style={
    draw=gray!70,
    line width=0.9pt,
    dash pattern=on 4pt off 4pt
  },
  Lcurve/.style={draw=red!80!black, line width=1.05pt},
  zpoint/.style={
    circle,
    draw=black,
    fill=white,
    line width=0.9pt,
    inner sep=2.4pt
  },
  Ppoint/.style={
    circle,
    draw=black,
    fill=black,
    inner sep=2.5pt
  }
}

% ---------------- Outer polygon ----------------
\coordinate (A) at (-4.15,  1.28);
\coordinate (B) at (-1.55,  3.15);
\coordinate (C) at ( 1.55,  3.15);
\coordinate (D) at ( 4.10,  1.22);
\coordinate (E) at ( 4.10, -1.40);
\coordinate (F) at ( 1.45, -3.02);
\coordinate (G) at (-2.00, -3.02);
\coordinate (H) at (-4.15, -1.35);

\draw[boundary] (B) -- node[midway, above left=5pt] {$a_2$} (A);
\draw[boundary] (C) -- node[midway, above=7pt] {$b_2$} (B);
\draw[boundary] (C) -- node[midway, above right=5pt] {$a_1^{-1}$} (D);
\draw[boundary] (D) -- node[midway, right=10pt] {$b_1^{-1}$} (E);
\draw[boundary] (F) -- node[midway, below right=5pt] {$a_1$} (E);
\draw[boundary] (G) -- node[midway, below=8pt] {$b_1$} (F);
\draw[boundary] (G) -- node[midway, below left=5pt] {$a_2^{-1}$} (H);
\draw[boundary] (H) -- node[midway, left=10pt] {$b_2^{-1}$} (A);

% ---------------- Interior points ----------------
\coordinate (P)  at ( 0.00,-1.42);
\coordinate (z1) at (-1.10, 1.05);
\coordinate (z2) at ( 1.38, 0.55);
\coordinate (z3) at (-2.35,-0.32);

% ---------------- One connected keyhole loop L ----------------
% Straight path parts + open circular arcs near the z_i.
\draw[Lcurve]
  ( 0.23469,-1.23313) -- ( 1.28455, 0.26559)
  arc[start angle=251.45, end angle=578.53, radius=0.30]
  -- ( 0.09545,-1.13559)
  arc[start angle= 71.45, end angle= 97.55, radius=0.30]
  -- (-0.90531, 0.82176)
  arc[start angle=310.46, end angle=637.55, radius=0.30]
  -- (-0.19469,-1.19176)
  arc[start angle=130.46, end angle=138.46, radius=0.30]
  -- (-2.05339,-0.36499)
  arc[start angle=351.38, end angle=678.46, radius=0.30]
  -- (-0.29661,-1.37501)
  arc[start angle=171.38, end angle=398.53, radius=0.30];

% ---------------- Dashed center paths ----------------
\draw[centerpath]
  (P) --
  node[pos=0.52, xshift=-17pt, yshift=2pt, text=red!80!black] {$\ell_1$}
  (z1);

\draw[centerpath] (P) -- (z2);
\draw[centerpath] (P) -- (z3);

% ---------------- Marked points and labels ----------------
\node[zpoint, label={[xshift=7pt,yshift=5pt]$z_1$}] at (z1) {};
\node[zpoint, label={[xshift=7pt,yshift=5pt]$z_2$}] at (z2) {};
\node[zpoint, label={[xshift=-9pt,yshift=6pt]$z_3$}] at (z3) {};

\node[Ppoint] at (P) {};
\node at (0.42,-1.69) {$P$};

\node[text=red!80!black] at (-1.88,1.33) {$h_1$};
\node[text=red!80!black] at (1.28,-0.10) {$L$};

\end{tikzpicture}%
}
\caption{Riemann surface of genus 2 with 3 marked points. }
\label{fig:sub1}
\end{figure}

When poles are present, one must supplement the usual bilinear identity by boundary contributions around the singularities. To make things clear, note that one can construct a Riemann surface by gluing the edges of a polygon \cite{griffiths2014principles, Eynard:2018tcr}. For example, a genus-2 Riemann surface is shown in Figure \ref{fig:sub1}. Let $\Sigma_0$ be the open polygon
\begin{equation}
    \Sigma_0 = \Sigma \setminus \cup_i(A_i\cup B_i).
\end{equation}
The $a_i$ and $a_i^{-1}$ sides of $\Sigma_0$ combine into the $A_i$ cycle. The dual cycle is $B_i$ coming from $b_i$'s. 
Let $U$ be a tubular neighborhood of $\p \Sigma_0$ in $\Sigma_0$. Within $U$, the following function is well-defined:
\begin{equation}
    f_1 (z) = \int_P^z dx~ \omega_1 (x),\qquad z \in U,
\end{equation}
where $P \in U$ is some fixed reference point. This is because on compact Riemann surfaces, $\sum \operatorname{Res} \omega_1 = 0$ for any meromorphic 1-form $\omega_1$. 

Now, suppose the singularities of $\omega_{1,2}$ are $\{z_i\}_{i=1}^N \subset \Sigma_0-U$. We can dig a hole $B_\epsilon(z_i)$ near each $z_i$. The function $f_1(z)$ is well-defined outside the keyhole loop $L$ as shown in the figure. We note that for $q \in a_1^{-1}$ and $q'\in a_1$ that are identified by the gluing map, we have
\begin{equation}
    f_1(q) - f_1(q') = - \oint_{B_1} \omega_1.
\end{equation}
Using this fact, we have
\begin{equation}
    \oint_{\p \Sigma_0} f_1 \bar \omega_2 = - \oint_{\p \Sigma_0} \bar f_2  \omega_1  =  \sum_{j=1}^g \z( \oint_{A_j} \omega_1 \oint_{B_j} \bar\omega_2  - \oint_{B_j} \omega_1 \oint_{A_j} \bar\omega_2 \y).
\end{equation}
If $\omega_{1},~\omega_2$ are holomorphic, then this is nothing but \eqref{holoRBI}. We now include the contributions from the poles. The regularization scheme we will take is to dig holes at the singularities, and link the holes using a keyhole region with boundary $L$. The integral we will compute is defined as the regularized integral:
\begin{equation}
    I =  \int_{\Sigma-\cup_{i}B_{\epsilon}(z_i)} \omega_1 \wedge \bar \omega_2 =  \int_{\Sigma_0-\operatorname{Int}(L)} \omega_1 \wedge \bar \omega_2,
\end{equation}
where $\operatorname{Int}(L)$ is the interior of $L$, namely the keyhole region. Thus, besides the contribution from the large cycles, we have another contribution
\begin{equation}
    I = \sum_{j=1}^g \z( \oint_{A_j} \omega_1 \oint_{B_j} \bar\omega_2  - \oint_{B_j} \omega_1 \oint_{A_j} \bar\omega_2 \y) - \oint_L f_1 \bar \omega_2.
\end{equation}
As an integral on a classical geometrical object, $I$ is well-defined if the sets of singularities of $\omega_1$ and $\bar \omega_2$ have an empty intersection, and there are only simple poles. That is, if $z_i$ is a simple pole of $\omega_1$, then it cannot be a pole of $\bar \omega_2$. Let us consider this situation in this subsection. 

Let us decompose $L = \cup_{i}(h_i \cup \ell_i )$, where $h_i=\p B_\epsilon(z_i)$ is the small circle surrounding $z_i$, $\ell_i$ is the difference of the paths connecting $P$ and $z_i$. On $\ell_i$, we have
\begin{equation}
    \int_{\ell_i} f_1 \bar{\omega}_2 = -2\pi i \operatorname{Res}_{z_i}(\omega_1) \int_{P}^{\bar z_i} \bar{\omega}_2,
\end{equation}
where the integral of $\bar \omega_2$ is along the right side of $\ell_i$. There would be a log divergence $\log(\epsilon)$ if $z_i$ is a simple pole of $\bar \omega_2$; however, by our assumption, that means $\operatorname{Res}_{z_i}(\omega_1) = 0$. On $h_i$, we have $f_1 \simeq \operatorname{Res}_{z_i}(\omega_1) \log(z-z_i)$. If $z_i$ is not a singularity of $\omega_2$, then we will just get zero for the integral on $h_i$. If $z_1$ is not a singularity of $\omega_1$, but a singularity of $\omega_2$, then we will get
\begin{equation}
    \int_{h_i} f_1 \bar{\omega}_2 = f_1(z_i) \oint_{\bar z_i} \bar \omega_2 = -2\pi i f_1(z_i) \operatorname{Res}_{\bar z_i}(\bar \omega_2) = -2\pi i \operatorname{Res}_{\bar z_i}(\bar \omega_2) \int_{P}^{ z_i} {\omega}_1.
\end{equation}
Here, for anti-meromorphic 1-forms, we define
\begin{equation}
  \bar  \omega_2 = \frac{A}{\bar z - \bar z_i} + \mathcal{O}(1) ,\quad \operatorname{Res}_{\bar z_i}(\bar \omega_2) = A.
\end{equation}
Thus, the classical RBI for meromorphic $\omega_1$ and anti-meromorphic $\bar \omega_2$ with at most simple poles and empty intersection of sets of poles is
\begin{equation}\label{RBI2.14}
\begin{aligned}
    I & = \sum_{j=1}^g \z( \oint_{A_j} \omega_1 \oint_{B_j} \bar\omega_2  - \oint_{B_j} \omega_1 \oint_{A_j} \bar\omega_2 \y) \\
    & \qquad \qquad  + \sum_{k=1}^N 2\pi i\z(   \operatorname{Res}_{z_k}(\omega_1) \int_P^{\bar{z}_k} \bar \omega_2 + \operatorname{Res}_{\bar z_k} (\bar \omega_2)\int_P^{{z}_k} \omega_1 \y). 
\end{aligned}
\end{equation}
It is helpful to introduce additional $A$-cycles $A_{g+i}=h_i$ and $B$-cycles $B_{g+i}= \operatorname{Path}(P,z_i)$, so that we have at total $(g+N)$ pairs of cycles. Then the formula can be rewritten as
\begin{equation}\label{classicalRBI}
    I = \sum_{j=1}^{g+N} \z( \oint_{A_j} \omega_1 \int_{B_j} \bar\omega_2  - \int_{B_j} \omega_1 \oint_{A_j} \bar\omega_2 \y). 
\end{equation}

\subsection{$J\bar J$-deformation as a Riemann bilinear identity}
Now, let us take a closer look at the first-order perturbation theory of $J\bar J$-deformation. Let $\mathcal O(w)$ be a local field of the infinitesimally deformed theory. We will encounter the following OPE in the correlation function:
\begin{equation}
    (J_1  \bar J_2) (z) \mathcal O (w) ~\sim~ : J_1(z)  \operatorname{OPE} (\bar J_2 \mathcal O)(w) : + : \operatorname{OPE} (J_1 \mathcal O)(w) \bar J_2(\bar z):+ \text{double-contractions}.
\end{equation}
Here, the normal ordering means there is no singularity in the limit $z \to w$ for this term. The situation can be understood more clearly if we also consider the correlation function
\begin{equation}
    \int d^2 z  ~ \langle J_1(z)\bar J_2(\bar z) \mathcal{O}_1(w_1) \cd \mathcal{O}_N(w_N) \rangle.
\end{equation}
The Ward identity of $\bp J_1(z)$ tells us that
\begin{equation}\label{Ward2.17}
    \langle (\bp J_1)(z) \mathcal{O}_1(w_1) \cd \mathcal{O}_N(w_N) \rangle = \text{contact terms}.
\end{equation}
Suppose the $J_1 \mathcal O_i$ OPE takes the form
\begin{equation}\label{J1OiOPE}
    J_1(z) \mathcal{O}_i (w_i) = \frac{q_i \mathcal{O}_i(w)}{z-w_i} + \sum_{n\geq 2} \frac{\mathcal{O}_{i,n}(w_i)}{(z-w_i)^n} + :J_1(z)\mathcal{O}_i (w_i):
\end{equation}
Then we have on the plane
\begin{equation}
\begin{aligned}
    \langle J_1(z) \mathcal{O}_1(w_1) \cd \mathcal{O}_N(w_N) \rangle &= \sum_{j=1}^{N} \sum_{n\geq 1} \langle \frac{\mathcal{O}_{j,n}(w_j)}{(z-w_j)^n} \mathcal O_{1}(w_1) \cd \widehat{\mathcal{O}_j(w_j)} \cd \mathcal O_{N}(w_N) \rangle  \\
    & \qquad \qquad \qquad + \langle J_1^{\mathrm{reg}}(z) \mathcal{O}_1(w_1) \cd \mathcal{O}_N(w_N) \rangle.
\end{aligned}
\end{equation}
Here $\widehat{\mathcal O_j(w_j)}$ means to delete this factor from the product, $\mathcal{O}_{j,1}(w)=q_j \mathcal{O}_j(w)$, and $J_1^{\mathrm{reg}}(z)$ is regular in the limit $z\to w_j$ for any $j$, thus capturing the regular part that is killed by $\bp$ in the Ward identity \eqref{Ward2.17}. Note that on a higher genus Riemann surface, the above equation can also be generalized, with the contact terms replaced by some globally well-defined meromorphic 1-form\footnote{The existence of such a meromorphic 1-form is guaranteed by the vanishing sum of residues, which is equivalent to charge conservation. } and the regular term by a globally well-defined holomorphic 1-form. We will see examples, e.g. \eqref{J14.45}, when we discuss the deformation of partition functions. For $J_1(z)\bar J_2(\bar z)$ insertion, it is useful to separate the result into off-diagonal contractions, contractions with the regular part of one current, diagonal double contractions, and the fully regular term:
\begin{equation}\label{J1J2OOO}
\begin{aligned}
    &\langle J_1(z) \bar J_2(\bar z) \mathcal{O}_1(w_1) \cd \mathcal{O}_N(w_N) \rangle \\
    =& \sum_{j\neq k} \sum_{n_1,n_2\geq 1} \langle \frac{\mathcal{O}_{j,n_1}(w_j)\overline{\mathcal{O}}_{k,n_2}(w_k)}{(z-w_j)^{n_1}(\bar z-\bar w_k)^{n_2}} \mathcal O_{1}(w_1) \cd \widehat{\mathcal{O}_j(w_j)} \cd \widehat{\mathcal{O}_k(w_k)} \cd  \mathcal O_{N}(w_N) \rangle \\
    & \qquad \qquad \qquad + \langle J_1^{\mathrm{reg}}(z) \sum_{k=1}^N \sum_{n_2\geq 0} \frac{\overline{\mathcal O}_{k,n_2}}{(\bar z - \bar w_k)^{n_2}} \mathcal{O}_1(w_1) \cd \widehat{\mathcal{O}_k(w_k)} \cd \mathcal{O}_N(w_N) \rangle \\
    & \qquad \qquad \qquad + \langle \bar J_2^{\mathrm{reg}}(\bar z) \sum_{j=1}^N \sum_{n_1\geq 0} \frac{{\mathcal O}_{j,n_1}}{( z -  w_j)^{n_1}} \mathcal{O}_1(w_1) \cd \widehat{\mathcal{O}_j(w_j)} \cd \mathcal{O}_N(w_N) \rangle \\
    &  \qquad \qquad \qquad + \text{diagonal double contractions}\\ 
    & \qquad \qquad \qquad + \langle J_1^{\mathrm{reg}}(z)\bar J_2^{\mathrm{reg}}(\bar z) \mathcal{O}_1(w_1) \cd \mathcal{O}_N(w_N) \rangle.
\end{aligned}
\end{equation}
The analysis becomes significantly simpler if there is only a simple pole in the OPE \eqref{J1OiOPE}. To expose the mechanism, we first assume that the current-OPEs with the inserted operators contain only simple poles. Then, \eqref{J1J2OOO} is simplified
\begin{equation}\label{simp2.21}
\begin{aligned}
    &\langle J_1(z) \bar J_2(\bar z) \mathcal{O}_1(w_1) \cd \mathcal{O}_N(w_N) \rangle \\
    =& \sum_{j\neq k} \langle \frac{q_j \bar q_k }{(z-w_j)(\bar z-\bar w_k)} \mathcal O_{1}(w_1) \cd  \mathcal O_{N}(w_N) \rangle \\
    & \qquad \qquad  + \langle J_1^{\mathrm{reg}}(z) \sum_{k=1}^N \frac{\bar q_k }{(\bar z - \bar w_k)} \mathcal{O}_1(w_1)  \cd \mathcal{O}_N(w_N) \rangle \\
    & \qquad \qquad  + \langle \bar J_2^{\mathrm{reg}}(\bar z) \sum_{j=1}^N  \frac{q_j}{( z -  w_j)} \mathcal{O}_1(w_1)  \cd \mathcal{O}_N(w_N) \rangle \\
    & \qquad \qquad  + \langle J_1^{\mathrm{reg}}(z)\bar J_2^{\mathrm{reg}}(\bar z) \mathcal{O}_1(w_1) \cd \mathcal{O}_N(w_N) \rangle \\
    &  \qquad \qquad  +\sum_{j=1}^N \langle \frac{q_j \bar q_j }{(z-w_j)(\bar z-\bar w_j)} \mathcal O_{1}(w_1)   \cd  \mathcal O_{N}(w_N) \rangle. \\
\end{aligned}
\end{equation}
Here, we still separate the diagonal double contraction terms out in the last line. The first four lines in the RHS above satisfy the conditions so that the classical RBI \eqref{classicalRBI} can be applied\footnote{One can also use conformal perturbation theory, giving the same result. }. There will be some log-cuts in the RBI of the off-diagonal double contraction in the first term of the RHS in \eqref{simp2.21}. To make the structure clearer, it is helpful to consider the ``dressing\footnote{Note that the notion of dressed operators is not the same one defined in \cite{Kruthoff:2020hsi}. }" of a single operator:
\begin{equation}
\begin{aligned}
    &\int_{\mathbb C - \mathrm{Path}(P,w)} J_1(z)dz\wedge \bar{J}_2(\bar z)d\bar z ~ \mathcal O(w) \\
 = & 2\pi i q  :\mathcal O(w) \int_P^{\bar w}  d\bar z~\bar J_2(\bar z): + 2\pi i \bar q : \mathcal O(w)\int_P^w  d  z~J_1( z) : \\
 & \qquad\qquad\qquad\qquad\qquad +\text{ double-contraction term } \propto ~ q\bar q \mathcal{O}(w),
\end{aligned}
\end{equation}
where the first two terms are direct applications of \eqref{classicalRBI} with $\omega_1 = J_1(z)dz$ and $\bar \omega_2=\bar J_2(\bar z) d\bar z$. The double-contraction term is divergent and proportional to $\mathcal{O}(w)$ itself. We introduce bosonization for the currents,
\begin{equation}\label{bosonization}
    \phi_1(w) = \int_P^w dz~ J_1(z) ,\qquad \bar \phi_2(\bar w) = \int_P^{\bar w} d\bar z~ \bar J_2(\bar z). 
\end{equation}
Then, the dressing becomes
\begin{equation}
\begin{aligned}
    \int_{\mathbb C - \mathrm{Path}(P,w)} J_1(z)dz\wedge \bar{J}_2(\bar z)d\bar z ~ \mathcal O(w) & = 2\pi i q  :\mathcal O(w) \bar \phi_2(\bar w) : +  2\pi i \bar q : \mathcal O(w) \phi_1(w) : \\
     & \qquad  +  \text{ double-contraction term } \propto ~ q\bar q \mathcal{O}(w).
\end{aligned}
\end{equation}
This means that the leading order dressing of $\mathcal{O}(w)$ is
\begin{equation}\label{dresssingle}
\begin{aligned}
    \mathcal{O}(w) & \mapsto \Dr_{\delta \lambda}(\mathcal{O})(w) = \mathcal{O}(w) + \delta \lambda \Big( \pi  q  :\mathcal O(w) \bar \phi_2(\bar w) : +  \pi  \bar q : \mathcal O(w) \phi_1(w) :  \Big) \\
    & \qquad\qquad \qquad\qquad\qquad\qquad \qquad + \text{ double-contraction term } \propto ~ q\bar q \mathcal{O}(w).
\end{aligned}
\end{equation}
We note that \eqref{dresssingle} is consistent with \eqref{simp2.21}, so that the log-terms from the off-diagonal double-contractions are contractions of the bosons. Indeed, the bosonic fields $\phi_1,\bar{\phi}_2$ as a field operator on the Riemann surface have singular OPEs with the fields $\mathcal{O}_i$. Although they are put in normal orderings with $\mathcal O(w)$ in \eqref{dresssingle}, they can still be contracted with other possible insertions. Thus, at the order $\mathcal{O}(\delta \lambda)$, we have
\begin{equation}\label{coranddress}
\begin{aligned}
   & \langle \z( 1 - \delta \lambda \int_{\mathbb C - \operatorname{Int} L} d^2 z~ J_1(z) \bar{J}_2(\bar z)  \y)  \mathcal O_{1}(w_1) \cd  \mathcal O_{N}(w_N)  \rangle_0 \\
   & \qquad \qquad \qquad \qquad \qquad \simeq  \langle  \Dr_{\delta \lambda}(\mathcal{O}_1)(w_1) \cd \Dr_{\delta \lambda}(\mathcal{O}_N)(w_N)  \rangle_0.
\end{aligned}
\end{equation}
This is not a surprise, since it is just using two different ways to calculate the same thing. Equation \eqref{coranddress} not only matches the regular parts but also matches the singular parts, which can be subtracted by  some subtraction scheme. On the LHS of \eqref{coranddress}, by our assumption we can use conformal perturbation theory and choose the renormalization scheme so that we have \eqref{eq1.4}. Then, we can accordingly renormalize the dressing so that the renormalized version of \eqref{eq2.5} is satisfied:
\begin{equation}\label{rencoranddress}
\begin{aligned}
   & \langle \z( 1 - \delta \lambda \int_{\mathbb C - \operatorname{Int} L} d^2 z~ J_1(z) \bar{J}_2(\bar z)  \y)  \mathcal O_{1}(w_1) \cd  \mathcal O_{N}(w_N)  \rangle_{0,\ren} \\
   & \qquad \qquad \qquad \qquad \qquad \simeq  \langle  \Dr^{\ren}_{\delta \lambda}(\mathcal{O}_1)(w_1) \cd \Dr^{\ren}_{\delta \lambda}(\mathcal{O}_N)(w_N)  \rangle_0 \\
   &  \qquad \qquad \qquad \qquad \qquad \simeq   \langle \mathcal O_{1}(w_1) \cd  \mathcal O_{N}(w_N)  \rangle_{\delta \lambda} ,\qquad \text{at $\mathcal O(\delta \lambda)$}.
\end{aligned}
\end{equation}
In the above equations, we can replace $(\mathbb C -\operatorname{Int} L )$ by $(\mathbb{C}P^1 -\operatorname{Int} L )$. This is because there is no anomaly in $J_1$ and $\bar J_2$, and the sum of the charges should vanish to get a non-zero correlation function, i.e., $\sum_i q_i = \sum_i \bar q_i = 0$. If we generalize the discussion to a higher genus Riemann surface, then we must take into account the contribution from the large-cycle part of the RBI, which is the deformation of the partition function part. We will discuss this deformation in detail in Section \ref{sec:partitionfunc}. However, for the deformation of local operators, the construction \eqref{dresssingle} should be valid in higher genus cases. The dressing of a local operator is still local; the non-trivial topology is not important when discussing the deformation of local operators. 

Note that the bare first-order RBI dressing can only be directly applied to the operator $\mathcal{O}(w)$ whose OPEs with $J_1(z)$ and $\bar J_2(\bar z)$ only contain simple poles. For higher order poles, there will be higher divergent terms containing negative power of the UV cutoff $\epsilon^{-n},~n>0$. However, the same divergences appear in the conformal perturbation theory as well. As we have assumed there is a regularization and renormalization scheme to take care of those divergent integrals in the conformal perturbation theory, one can use the same scheme to address the singular integrals in the dressing. Thus, the renormalized dressing map exists even for operators whose OPE with $J_1$ and $\bar J_2$ contains higher poles. In particular, the renormalized dressing of the current $J_1(z)$ and the stress tensor $T(z)$ exists. 

To be more precise, equation \eqref{eq2.5} suggests that the renormalized dressing should be a linear map:
\begin{equation}
\begin{aligned}
    \Dr^{\ren}_{\delta \lambda} : \mathcal{H}_{\delta\lambda} &\rightarrow \mathcal{H}_0,
\end{aligned}
\end{equation}
where $\mathcal{H}_{\delta\lambda}$ is the space of operators in the $S_{\delta \lambda}$-theory, while $\mathcal{H}_0$ is that of the undeformed theory. By our proposal \eqref{rencoranddress}, the correlation functions with current $J^{(\delta \lambda)}_1$ insertions should be preserved under dressing. In particular, the norm is preserved
\begin{equation}
\begin{aligned}
    J^{(\delta \lambda)}_1(z)J^{(\delta \lambda)}_1(w) &\sim - \frac{\kappa}{(z-w)^2} \quad\text{in $S_{\delta \lambda}$-theory} \\
    \Dr^{\ren}_{\delta \lambda}(J^{(\delta \lambda)}_1)(z)  \Dr^{\ren}_{\delta \lambda}(J^{(\delta \lambda)}_1)(w) & \sim - \frac{\kappa}{(z-w)^2} \quad\text{in $S_{0}$-theory}. 
\end{aligned}
\end{equation}
Since the current algebra is preserved, the dressed current must be identified with the normalized current of the undeformed theory:
\begin{equation}
    \Dr^{\ren}_{\delta \lambda}(J^{(\delta \lambda)}_1)(z) = J_1^{(0)}(z)  \quad\text{in $S_{0}$-theory}. 
\end{equation}
The same argument applied to the stress tensor gives
\begin{equation}
    \Dr^{\ren}_{\delta \lambda}(T^{(\delta \lambda)})(z) = T^{(0)}(z)  \quad\text{in $S_{0}$-theory}. 
\end{equation}
In particular, the central charge is invariant under the deformation. 

Since we assume a renormalized conformal perturbation theory exists for all orders, we can integrate the dressing correspondingly to claim
\begin{equation}\label{DrXXX2.35}
    \langle \Dr^\ren_\lambda(X^{(\lambda)}) \rangle_{0,\mathbb{C}P^1 } = \langle X^{(\lambda)} \rangle_{\lambda,\mathbb{C}P^1}.
\end{equation}
Note that for a genus $g\geq 1$ worldsheet, the RBI also contributes a large-cycle integral, and thus there will be such an additional term on the LHS. We will see examples and analyze them carefully in Section \ref{sec:partitionfunc}. After integration, the dressing should be understood as a linear map
\begin{equation}\label{rendress2.36}
\begin{aligned}
    \Dr^{\ren}_{\lambda} : \mathcal{H}_{\lambda} &\rightarrow \mathcal{H}_{0} \\
    J_1^{(\lambda)} & \to J_1^{(0)}\\
    T^{(\lambda)} & \to T^{(0)}.
\end{aligned}
\end{equation}
Note that if there is no self-interaction, the bare perturbative dressing on the current $J_1$ will be trivial. However, it does not imply that the renormalized dressing of $J_1$ is trivial. It just means that the diagonal entry of $J_1$ in the Zamolodchikov metric \cite{Zamolodchikov:1986gt} is zero; they must still be paired with some other fields in a unitary CFT. Then, the situation depends on whether the pairing takes place only in the $O(2,2)$ subsystem of bosonization of $J_1$ and $\bar J_2$. We will carefully analyze examples for both cases in Section \ref{lcexample1}.

\subsection*{Example: one free boson}\label{noncompactfreeboson}
As a simple example to illustrate the ideas, let us consider
a non-compact free boson
\begin{equation}\label{actionboson}
    S_t= \frac{t}{\pi} \int d^2 z ~\p X \bp X.
\end{equation}
The basic OPEs are
\begin{equation}\label{basicOPEfreeboson}
\begin{aligned}
    X(z)X(w) &\sim - \frac{1}{2t} \log |z-w|^2 \\
    e^{ipX(z)}e^{iq X(w)} &= |z-w|^{\frac{pq}{t}} e^{ipX(z)+iq X(w)} \\
    X(z) e^{ikX(w)} & \sim - \frac{1}{2t} ik \log |(z-w)/\Lambda|^2 e^{ikX(w)}.
\end{aligned}
\end{equation}
Now, let us consider the bare dressing on $e^{ikX}$. It is more convenient to use the non-flat coordinate $t$ to parametrize the deformation in this case:
\begin{equation}\label{2.3047}
\begin{aligned}
 &\z( \frac{\delta t}{\pi}  \int d^2z ~ \p X(z) \bp \bar X(\bar z) \y)  e^{ikX(w,\bar w)} =  \z( \frac{i\delta t}{2\pi}  \int \p X(z) dz \wedge \bp \bar X(\bar z) d\bar z \y)  e^{ikX(w,\bar w)} \\
 =&  \z( \frac{\delta t}{2t} \y) i k : (X(w)+\bar X(\bar w)) e^{ikX(w,\bar w)} : - \frac{\delta t}{2\pi} \frac{k^2}{4t^2} e^{ikX(w,\bar w)}  \int d^2 z \frac{1}{|z-w|^2}  \\
 = & \z( \frac{\delta t}{2t} \y) i k : X(w,\bar w) e^{ikX(w,\bar w)} : + \text{ singular double-contraction}. 
\end{aligned}
\end{equation}
This means the dressing will modify the momentum:
\begin{equation}
    k \mapsto \z( 1-\frac{\delta t}{2t} \y) k.
\end{equation}
As a comparison, it is also helpful to directly consider the deformation of correlation functions for comparison:
\begin{equation}
    \langle \prod_{i} e^{ik_i X(z_i)} \rangle = \int \mathcal{D}X ~e^{-S_0 -S_1}  \prod_{i} e^{ik_i X(z_i)}, 
\end{equation}
where
\begin{equation}
    \begin{aligned}
        S_0 &= \frac{t}{\pi} \int d^2 z ~\p X \bp X \\
        S_1 &= \frac{\delta t}{\pi} \int d^2 z ~\p X \bp X.
    \end{aligned}
\end{equation}
The result is clearly
\begin{equation}
    \langle \prod_{i} e^{ik_i X(z_i)} \rangle = \prod_{i<j} |z_{ij}|^{\frac{k_i k_j}{t+\delta t}}.
\end{equation}
Now, we want to expand $S_1$ to be a dressing on the vertices. After the dressing, we returned to the original $S_0$-theory. Then the correlation function should be:
\begin{equation}
    \prod_{i<j} |z_{ij}|^{\frac{k_i k_j}{t+\delta t}} = \prod_{i<j} |z_{ij}|^{\frac{k'_i k'_j}{t}}.
\end{equation}
This means
\begin{equation}\label{2.44}
    k'_i = \sqrt{\frac{t}{t+\delta t}} k_i \approx \z( 1 - \frac{\delta t}{2t} \y) k_i,
\end{equation}
which is consistent with \eqref{2.3047} up to normalization. The singular normalization in \eqref{2.3047} can be absorbed by introducing a field-strength renormalization with the renormalization condition in which the vertex has the same norm as the undressed operator \cite{Sen:2017gfr}. In this case, the renormalized dressing should be
\begin{equation}
    \Dr^\ren_{T\to t} (e^{ikX})(z,\bar z) = e^{i\sqrt{\frac{t}{T}}k X(z,\bar z)}.
\end{equation}
Note that the operator put in the dressing should always be understood as an operator in the deformed theory. Equivalently, we have
\begin{equation}\label{Xboson2.56}
     \Dr_{T\to t} (X(z,\bar z)) = \sqrt{\frac{t}{T}} X(z,\bar z) =  \Dr^\ren_{T\to t} (X(z,\bar z)).
\end{equation}
Now, if we normalize the current $J^{(t)}$ in the $S_t$-theory so that
\begin{equation}
    J^{(t)}(z)J^{(t)}(w) = -\frac{1}{2t} \frac{1}{(z-w)^2} \quad \text{in $S_t$-theory}.
\end{equation}
Then
\begin{equation}
    J^{(T)}(z)  = \sqrt{\frac{T}{t}} \p X \quad \text{in $S_T$-theory}.
\end{equation}
Then it is consistent with \eqref{Xboson2.56}:
\begin{equation}
    \Dr^\ren_{T\to t}(J^{(T)}) = J^{(t)}.
\end{equation}
As a consequence, the dressed deformed stress tensor is equal to the undeformed stress tensor. We can also recover the flat coordinate by
\begin{equation}
    \delta S_T = \frac{ t\delta T}{\pi T} \int d^2 z ~J^{(T)} \bar {J}^{(T)}. 
\end{equation}
Thus
\begin{equation}
    \frac{dT}{d\lambda} = \pi \frac{T}{t}.
\end{equation}
We can also consider compactifying the boson
\begin{equation}
    X(z,\bar z) \sim X(z,\bar z) + 2\pi R.
\end{equation}
Then, at the reference point $t$, the local vertex operators are parametrized by two integers $m,w$ as
\begin{equation}
    V_{m,w}^{(t)}(z,\bar z) = \exp \z( ip^{(t)} X(z) + i\bar{p}^{(t)} \bar X(\bar z) \y),\quad p^{(t)} = \frac{m}{R} + wR t ,\quad \bar p^{(t)} = \frac{m}{R} - wR t. 
\end{equation}
Now, if an operator in the $S_T$-theory is dressed back to the $S_t$-theory, the periodicity of the dressed field appears different:
\begin{equation}
    \begin{aligned}
        X(z,\bar z) & \sim X(z,\bar z) + 2\pi R \quad\text{in the $S_T$-theory} \\
        \Dr^\ren_{T \to t}( X)(z,\bar z) & \sim  \Dr^\ren_{T \to t}( X)(z,\bar z) + 2\pi R \quad\text{in the $S_t$-theory} \\
        \Rightarrow \quad X(z,\bar z) & \sim X(z,\bar z) + 2\pi R \sqrt{\frac{T}{t}} \quad\text{in the $S_t$-theory}. \\
    \end{aligned}
\end{equation}
This deformed boundary condition changes the spectrum of mutually local operators in the $S_t$-theory:
\begin{equation}
    \begin{aligned}
        V_{m,w}^{(T)}(z,\bar z) &= \exp \z( ip^{(T)} X(z) + i\bar{p}^{(T)} \bar X(\bar z) \y),\quad p^{(T)} = \frac{m}{R} + wR T ,\quad \bar p^{(T)} = \frac{m}{R} - wR T \\
       \Dr^\ren_{T \to t}(V_{m,w}^{(T)}) &= \exp \z( i \sqrt{\frac{t}{T}} p^{(T)} X(z) + i\sqrt{\frac{t}{T}} \bar{p}^{(T)} \bar X(\bar z) \y)  \quad \text{in $S_t$-theory}.
    \end{aligned}
\end{equation}
As operators in the $S_t$-theory, these dressed operators are mutually local with $X(z,\bar z)$ with the twisted boundary condition:
\begin{equation}
\begin{aligned}
   & X(z,\bar z) \Dr^\ren_{T \to t}(V_{m,w}^{(T)})(0,0)\\
   & \qquad\qquad \sim \z( - \frac{i}{2t} \sqrt{\frac{t}{T}}p^{(T)} \log(z) - \frac{i}{2t} \sqrt{\frac{t}{T}}\bar p^{(T)} \log(\bar z) \y)\Dr^\ren_{T \to t}(V_{m,w}^{(T)})(0,0).
\end{aligned}
\end{equation}
The mutual locality condition is
\begin{equation}
    \frac{\pi}{\sqrt{tT}} \z( p^{(T)} - \bar{p}^{(T)} \y) \in 2\pi R\sqrt{\frac{T}{t}} \mathbb Z.
\end{equation}
This is indeed the case. Thus, the dressing can also be understood as twisting the boundary conditions in the original undeformed theory, thereby changing the local spectrum. The conformal weight of the dressed operator is
\begin{equation}
    h\z( \Dr^\ren_{T \to t}(V_{m,w}^{(T)}) \y) = \frac{\z(\frac{t}{T} \z( \frac{m}{R} + wRT \y) \y)^2}{2t} = \frac{\z( \frac{m}{R} + wRT \y)^2}{2T} \quad \text{in $S_t$-theory}.
\end{equation}
This is exactly the same as the conformal weight of $V^{(T)}_{m,w}$ in the $S_T$-theory. 

Moreover, let us present the deformation of the partition function for later comparison:
\begin{equation}
\begin{aligned}
    Z_T(\tau,\bar\tau;\chi,\bar\chi)
    =
    \frac{1}{|\eta(\tau)|^2}
    \sum_{m,w\in\mathbb Z}
    \exp\Bigg[
        &-\pi\tau_2
        \left(
            \frac{m^2}{R^2T}+w^2R^2T
        \right)
        +2\pi i\tau_1mw +  \chi p^{(T)}+\bar\chi\bar p^{(T)} \Bigg].
\end{aligned}
\end{equation}
Here $\chi,\bar \chi$ are chemical potentials. The infinitesimal variation in the flat coordinate is
\begin{equation}
    \p_\lambda Z_\lambda = \frac{dT}{d\lambda} \frac{d}{dT}Z_T = \z( 4\pi^2 \tau_2 \p_\chi \p_{\bar \chi} - \frac{\pi}{2t} \chi \p_{\bar \chi} - \frac{\pi}{2t} \bar \chi \p_{ \chi}  \y) Z_\lambda.
\end{equation}
We have used the derivative of the charges
\begin{equation}
    \p_T p^{(T)} = - \frac{1}{2T} \bar p^{(T)} ,\qquad\p_T \bar p^{(T)} = - \frac{1}{2T} p^{(T)}. 
\end{equation}
These results are consistent with \eqref{3.55} and \eqref{difffffffffqh}. 

Note that although the bare dressing on $X(z,\bar z)$ is the renormalized one, it is not the case for the non-local field
\begin{equation}
    \Theta(z,\bar z) = X(z) -\bar {X}(\bar z).
\end{equation}
In fact, the bare dressing is in the reverse direction compared to \eqref{Xboson2.56}:
\begin{equation}
    \Dr_{T\to t}^{\bare} (\Theta) = \sqrt{\frac{T}{t}} \Theta \quad \neq  \quad \Dr_{T\to t}^\ren(\Theta) = \sqrt{\frac{t}{T}} \Theta.
\end{equation}
This problem also happens in the conformal perturbation theory calculation of the correlation functions involving $\Theta$. Thus, one should really include non-trivial renormalization so that we have \eqref{rencoranddress}. We also note that the bare dressing of $\Theta$ is the renormalized dressing in the $T$-dual frame $tR^2 \mapsto 1/(tR^2)$.

\section{Deformation of the partition functions}\label{sec:partitionfunc}

In this section, we study the deformation of partition functions with
chemical potentials. The discussion is simplest if there is no self-interaction, namely if $\kappa$ and $\bar\kappa$ vanish. We first discuss this case for the torus and higher genus partition functions. Then we include self-interactions. 

\subsection{No self-interaction cases}
If there is no self-interaction, the deformation can be directly expanded using the Riemann bilinear identity. However, as we have emphasized, renormalization may still generate a non-trivial dressing on the currents. Thus, we still need to choose $\lambda$ to be the flat coordinate. 

\subsubsection*{Torus partition function}
Consider the torus
\begin{equation}
    T_\tau=\mathbb C/(\mathbb Z+\tau\mathbb Z),
    \qquad
    \tau=\tau_1+i\tau_2,
    \qquad
    \tau_2>0.
\end{equation}
We call the spatial cycle the \(A\)-cycle and the time cycle the \(B\)-cycle.
We define the real normalized charges by
\begin{equation}\label{defchargeQQbar}
   Q^{(\lambda)}=\frac{1}{2\pi}\oint_A J^{(\lambda)}_1(z)dz,
    \qquad
    \bar Q^{(\lambda)}=-\frac{1}{2\pi}
    \oint_A\bar J^{(\lambda)}_2(\bar z)d\bar z. 
\end{equation}
The minus sign in the definition of \(\bar Q\) follows from the anti-holomorphic residue convention
\begin{equation}
    \oint\frac{d\bar z}{\bar z-\bar w}=-2\pi i .
\end{equation}
The torus partition function with chemical potentials is
\begin{equation}\label{z0tchi3.4}
    Z_0(\tau,\bar\tau;\chi,\bar\chi)
    =
    \left\langle
        \exp(\chi Q^{(0)}+\bar\chi\bar Q^{(0)})
    \right\rangle_0.
\end{equation}
The deformed partition function is
\begin{equation}
    Z_\lambda(\tau,\bar\tau;\chi,\bar\chi)
    =
    \langle
        \mathcal \exp(\chi Q^{(\lambda)}+\bar\chi\bar Q^{(\lambda)})
   \rangle_\lambda.
\end{equation}
It is convenient to consider the infinitesimal variation
\begin{equation}
\begin{aligned}
    \delta_\lambda Z_\lambda &= \langle e^{\chi Q(\lambda+\delta \lambda)+\bar \chi \bar Q(\lambda + \delta \lambda)} \rangle_{\lambda + \delta \lambda} -  \langle
         e^{\chi Q^{(\lambda)}+\bar\chi\bar Q^{(\lambda)}}
   \rangle_\lambda \\
   &= \langle e^{\chi Q(\lambda+\delta \lambda)+\bar \chi \bar Q(\lambda + \delta \lambda)} e^{-\delta \lambda \int d^2 z ~J_1^{(\lambda)} \bar J_2^{(\lambda)}}\rangle_{\lambda} -  \langle
         e^{\chi Q^{(\lambda)}+\bar\chi\bar Q^{(\lambda)}}
   \rangle_\lambda \\
   &= \langle \Dr^\ren_{\delta \lambda}( e^{\chi Q(\lambda+\delta \lambda)+\bar \chi \bar Q(\lambda + \delta \lambda)})\rangle_{\lambda}   -  \langle
         e^{\chi Q^{(\lambda)}+\bar\chi\bar Q^{(\lambda)}}
   \rangle_\lambda \\
   & \qquad -\delta \lambda\Big\langle e^{\chi Q(\lambda+\delta \lambda)+\bar \chi \bar Q(\lambda + \delta \lambda)} \int_{\text{large-cycle}} d^2z ~J^{(\lambda)}_1(z)\bar J^{(\lambda)}_2(\bar z) \Big\rangle_{\lambda},
\end{aligned}
\end{equation}
where we convert the deformation area integral into a RBI on large $A,B$-cycles and a renormalized dressing on the operators, just as \eqref{RBI2.14}. The net renormalized dressing turns the currents in $e^{\chi Q^{(\lambda+\delta \lambda)}+\bar \chi \bar Q^{(\lambda+\delta \lambda)}}$ back to the normalized currents in the $S_\lambda$-theory:
\begin{equation}
\begin{aligned}
    &\Dr_{\delta\lambda}^{\ren}(J_1^{(\lambda+\delta \lambda)}) = J_1^{(\lambda)} ,\qquad \Dr_{\delta \lambda}^{\ren}(\bar J_2^{(\lambda)}) = \bar J_2^{(\lambda)}\\
    &\Rightarrow \quad \Dr_{\delta \lambda}^{\ren}(e^{\chi Q^{(\lambda+ \delta \lambda)}+\bar \chi \bar Q^{(\lambda+\delta \lambda)}}) = e^{\chi Q^{(\lambda)}+\bar \chi \bar Q^{(\lambda)}}.
\end{aligned}
\end{equation}
Thus, the only contribution comes from large cycles:
\begin{equation}
\begin{aligned}
    \p_\lambda Z_\lambda(\tau,\bar\tau;\chi,\bar\chi) &= - \Big\langle e^{\chi Q(\lambda)+\bar \chi \bar Q(\lambda)} \int_{\text{large-cycle}} d^2z ~J^{(\lambda)}_1(z)\bar J^{(\lambda)}_2(\bar z) \Big\rangle_{\lambda} \\
    &=- \frac{i}{2} \Big\langle e^{\chi Q(\lambda)+\bar \chi \bar Q(\lambda)} \left(
        \oint_AJ^{(\lambda)}_1\oint_B\bar J^{(\lambda)}_2
        -
        \oint_BJ^{(\lambda)}_1\oint_A\bar J^{(\lambda)}_2
    \right) \Big\rangle_{\lambda}.
\end{aligned}
\end{equation}
Expanding in the chemical potentials reduces the problem to cycle integrals of current correlators of the form
\begin{equation}
   \oint_{A,B} dz  \langle J^{(\lambda)}_1(z) J^{(\lambda)}_1(z_1) \cd J^{(\lambda)}_1(z_N) \cd   \rangle_\lambda.
\end{equation}
Since there is no self-interaction, Ward identity indicates that $J_1(z)dz$ should be an operator-valued holomorphic 1-form on $T^2$. There is only one such holomorphic 1-form on $T^2$, namely $dz$ itself, we must have comparing with \eqref{defchargeQQbar}
\begin{equation}
    \oint_BJ^{(\lambda)}_1
    =
    2\pi\tau Q^{(\lambda)} ,\qquad \oint_B\bar J^{(\lambda)}_2
    =
    -2\pi\bar\tau\bar Q^{(\lambda)}. 
\end{equation}
We find the large-cycle integral becomes
\begin{equation}
\begin{aligned}
    \frac{i}{2}
    \left(
        \oint_AJ^{(\lambda)}_1\oint_B\bar J^{(\lambda)}_2
        -
        \oint_BJ^{(\lambda)}_1\oint_A\bar J^{(\lambda)}_2
    \right)=
    -4\pi^2\tau_2 Q^{(\lambda)}\bar Q^{(\lambda)} = -4\pi^2 \tau_2 \p_\chi \p_{\bar \chi}. 
\end{aligned}
\end{equation}
Therefore, the flow equation is
\begin{equation}\label{flowZ3.12}
   \p_\lambda Z_\lambda(\tau,\bar\tau;\chi,\bar\chi) = 4\pi^2 \tau_2 \p_\chi \p_{\bar \chi} Z_\lambda(\tau,\bar\tau;\chi,\bar\chi).
\end{equation}
The integrated version is
\begin{equation}
    Z_\lambda(\tau,\bar\tau;\chi,\bar\chi)
    =
    \exp\left(
        4\pi^2\lambda\tau_2
        \p_\chi\p_{\bar\chi}
    \right)
    Z_0(\tau,\bar\tau;\chi,\bar\chi).
\end{equation}
Equivalently, as a formal Gaussian convolution,
\begin{equation}\label{kernel3.14}
\begin{aligned}
    Z_\lambda(\tau,\bar\tau;\chi,\bar\chi)
    =
    \int_{\mathbb C}
    \frac{d^2\xi}{4\pi^3\lambda\tau_2}
    \exp\left[
        -\frac{|\xi|^2}{4\pi^2\lambda\tau_2}
    \right]
    Z_0(\tau,\bar\tau;\chi+\xi,\bar\chi+\bar\xi).
\end{aligned}
\end{equation}
The Gaussian expression should be understood by analytic continuation when \(\lambda\) is not in the convergent region. The same smearing kernel and the deformed spectrum discussed later have been found in \cite{Chakraborty:2024mls, Hashimoto:2019wct}. 

\subsubsection*{Higher genus}
One of the advantages of introducing the Riemann bilinear identity is that the method can be easily generalized to higher genus cases. Consider a genus \(g\) Riemann surface \(\Sigma_g\), and choose a
canonical homology basis \(\{A_I,B_I\}_{I=1}^g\) so that $A_I \cap B_I = \delta_{IJ}$. On $\Sigma_g$, there are $g$ holomorphic 1-forms $\omega_I$. We normalize them as
\begin{equation}\label{normalomega4.14}
    \oint_{A_I}\omega_J=\delta_{IJ},
    \qquad
    \oint_{B_I}\omega_J=\Omega_{IJ}.
\end{equation}
The matrix $\Omega_{IJ}$ is called the period matrix, which is symmetric $\Omega_{IJ} = \Omega_{JI}$ and encodes the moduli of the complex structure. Let $M = \operatorname{Im} \Omega$, then $M$ is positive-definite. We have
\begin{equation}\label{omegaomegaM}
\begin{aligned}
    \frac{i}{2} \int_{\Sigma_g} \omega_K \wedge \bar{\omega}_J &= \frac{i}{2} \sum_{I=1}^g \z(  \oint_{A_I} \omega_K \oint_{B_I} \bar{\omega}_J - \oint_{B_I} \omega_K \oint_{A_I} \bar{\omega}_J  \y) \\
    & = \frac{i}{2} \sum_{I=1}^g \z(  \delta_{IK} \bar{\Omega}_{IJ} - \Omega_{IK} \delta_{IJ} \y) \\
    & =  \frac{i}{2} \sum_{I=1}^g \z(  \delta_{IK} \bar{\Omega}_{IJ} - \Omega_{IK} \delta_{IJ} \y) \\
    & = M_{KJ}.
\end{aligned}
\end{equation}
Define the normalized charges by the $A$-cycle integrals
\begin{equation}
    Q^{(\lambda)}_I=\frac{1}{2\pi}\oint_{A_I}J^{(\lambda)}_1,
    \qquad
    \bar Q^{(\lambda)}_I=-\frac{1}{2\pi}\oint_{A_I}\bar J^{(\lambda)}_2.
\end{equation}
We can also generalize the chemical potential as follows. Define
\begin{equation}
    \rho_I = (M^{-1})_{IJ} \bar{\omega}_J , \quad \bar{\rho}_I = (M^{-1})_{IJ} {\omega}_J. 
\end{equation}
Then
\begin{equation}
    \frac{i}{2} \int_{\Sigma_g} \omega_K \wedge \rho_I = \delta_{KI} ,\quad \frac{i}{2} \int_{\Sigma_g} \bar \rho_I \wedge \bar{\omega}_K = \delta_{IK}.
\end{equation}
Introduce the flat background source
\begin{equation}
    A^{0,1} (\chi ) = \frac{1}{2\pi} \chi_I \rho_I ,\quad A^{1,0} (\bar \chi ) =-\frac{1}{2\pi} \bar \chi_I \bar \rho_I.
\end{equation}
Then the chemical potential can be defined as
\begin{equation}
    \frac{i}{2} \int_{\Sigma_g}  J^{(\lambda)}_1(z) \wedge A^{0,1} (\chi) dz = \chi_I Q^{(\lambda)}_I ,\quad  \frac{i}{2} \int_{\Sigma_g} A^{1,0} (\bar \chi) \wedge \bar J^{(\lambda)}_2(\bar z) d\bar z = \bar \chi_I \bar Q_I ^{(\lambda)}.
\end{equation}
When there is no self-interaction, the currents are holomorphic and
anti-holomorphic one-forms, hence
\begin{equation}
    J^{(\lambda)}_1(z)dz=2\pi Q^{(\lambda)}_I\,\omega_I(z),
    \qquad
    \bar J^{(\lambda)}_2(\bar z)d\bar z=-2\pi\bar Q^{(\lambda)}_I\,\bar\omega_I(\bar z).
\end{equation}
Using \eqref{omegaomegaM}, we obtain 
\begin{equation}
\begin{aligned}
    S_1
    &=
    \int_{\Sigma_g}d^2z\,J^{(\lambda)}_1\bar J^{(\lambda)}_2
    =
    \frac{i}{2}\int_{\Sigma_g}
    J^{(\lambda)}_1(z)dz\wedge\bar J^{(\lambda)}_2(\bar z)d\bar z
    \\
    &=
    -4\pi^2 (Q^{(\lambda)})^T M\bar Q ^{(\lambda)}.
\end{aligned}
\end{equation}
Therefore
\begin{equation}
    Z_\lambda^{(g)}(\Omega;\chi,\bar\chi)
    =
    \exp\left(
        4\pi^2\lambda\,
        \p_\chi^T M \p_{\bar\chi}
    \right)
    Z_0^{(g)}(\Omega;\chi,\bar\chi).
\end{equation}
Equivalently,
\begin{equation}
\begin{aligned}
    Z_\lambda^{(g)}(\Omega;\chi,\bar\chi)
    =
    \int_{\mathbb C^g}
    \frac{d^{2g}\xi}
    {\pi^g\det(4\pi^2\lambda M)}
    \exp\left[
        -\xi^\dagger(4\pi^2\lambda M)^{-1}\xi
    \right]
    Z_0^{(g)}(\Omega;\chi+\xi,\bar\chi+\bar\xi).
\end{aligned}
\end{equation}

\subsubsection*{Modular invariance}
Let us discuss the modular invariance property of the formulae in the previous subsections. Again, it is helpful to consider the $g=1$ case. Consider the modular transformation
\begin{equation}\label{modulartorustransf}
    \tau' = \frac{a\tau + b}{c\tau +d} ,\quad \chi' = \frac{\chi}{c\tau + d} ,\quad \bar \chi' = \frac{\bar \chi}{c\bar \tau + d},\quad \tau_2' = \frac{\tau_2}{|c\tau + d|^2}, \quad \begin{pmatrix}
        a & b\\
        c & d
    \end{pmatrix} \in \mathrm{SL}(2,\mathbb Z).
\end{equation}
Assuming the undeformed theory is modular invariant
\begin{equation}
    Z_0 (\tau' , \bar \tau' ; \chi' , \bar \chi') = Z_0 (\tau , \bar \tau ; \chi , \bar \chi). 
\end{equation}
Since
\begin{equation}
    \p_{\chi'}=(c\tau+d)\p_\chi,
    \qquad
    \p_{\bar\chi'}=(c\bar\tau+d)\p_{\bar\chi},
\end{equation}
we have
\begin{equation}
    \tau_2'\p_{\chi'}\p_{\bar\chi'}
    =
    \tau_2\p_\chi\p_{\bar\chi}.
\end{equation}
The same invariance is visible directly at the level of the kernel:
\begin{equation}
    \frac{d^2 \chi' }{\pi \lambda \tau_2'} = \frac{d^2 \chi }{\pi \lambda \tau_2} ,\quad e^{-|\chi'-\nu'|^2/(\lambda \tau_2')} = e^{-|\chi-\nu|^2/(\lambda \tau_2)}.
\end{equation}
Thus, the deformation kernel preserves modular invariance if the undeformed partition function is modular invariant.

For higher genus, the modular group is \(\operatorname{Sp}(2g,\mathbb Z)\): 
\begin{equation}
    \Omega'=(A\Omega+B)(C\Omega+D)^{-1}.
\end{equation}
Here, the integral $g\times g$ matrices $A,B,C,D$ satisfying
\begin{equation}
    A^T D - C^T B = I_g, \quad A^TC = C^T A, \quad B^T D= D^T B.
\end{equation}
One has
\begin{equation}
    M'
    =
    (C\bar\Omega+D)^{-T}M(C\Omega+D)^{-1}.
\end{equation}
The source variables transform as
\begin{equation}
    \chi'=\chi(C\Omega+D)^{-1},
    \qquad
    \bar\chi'=\bar\chi(C\bar\Omega+D)^{-1}.
\end{equation}
Consequently
\begin{equation}
    \p_{\chi'}^T M'\p_{\bar\chi'}
    =
    \p_\chi^T M\p_{\bar\chi},
\end{equation}
so the higher-genus kernel is also modular invariant.

\subsection{Adding self-interactions}
\subsubsection*{Torus partition function}
We now allow the currents to have self-OPE double poles,
\begin{equation}\label{selfJJope}
    J^{(\lambda)}_1(z)J^{(\lambda)}_1(w)\sim -\frac{\kappa}{(z-w)^2},
    \qquad
    \bar J^{(\lambda)}_2(\bar z)\bar J^{(\lambda)}_2(\bar w)\sim
    -\frac{\bar\kappa}{(\bar z-\bar w)^2}.
\end{equation}
Thus, the Ward identity indicates that in
\begin{equation}\label{JdzJJJ}
   \langle J^{(\lambda)}_1(z) dz~  J^{(\lambda)}_1(z_1) \cd J^{(\lambda)}_1(z_N) \cd   \rangle_\lambda
\end{equation}
$J^{(\lambda)}_1(z)dz$ is a meromorphic 1-form with double poles at $z=z_i$, $i=1,2\cd,N$. For this case, we need a variant of the Riemann bilinear identity. Consider a meromorphic 1-form $\omega$ with poles at $z_i$. As in the discussion in Section \ref{secgeoRBI}, there is a primitive of $\omega$ on $\Sigma_0-\operatorname{Int}L$
\begin{equation}
    f (z) = \int_P^z \omega,\qquad z \in \Sigma_0-\operatorname{Int}L.
\end{equation}
We can consider the trivial integral
\begin{equation}
\begin{aligned}
    0 &= \int_{\Sigma_0 - \operatorname{Int}L} \omega \wedge dz =  \oint_{A} \omega \oint_{B} dz - \oint_{B} \omega \oint_{A} dz  - \oint_L f dz \\
    &= \tau \oint_{A} \omega  - \oint_{B} \omega   - \oint_L f dz. 
\end{aligned}
\end{equation}
In particular, if $\omega$ has vanishing residues at all the poles
\begin{equation}
    \omega(z) = \frac{a_idz}{(z-z_i)^2} +  \frac{b_idz}{(z-z_i)^3} + \cd,
\end{equation}
then, the primitive is locally
\begin{equation}
    f(z) = - \frac{a_i}{z-z_i} + \cd. 
\end{equation}
Thus
\begin{equation}\label{omega4.47}
    \oint_{B} \omega  - \tau \oint_{A} \omega = 2\pi i \sum_i a_i.
\end{equation}
In particular, $\omega$ has a global meromorphic primitive $f$ if and only if $\sum_i a_i = 0$. This condition is nothing but $\sum \operatorname{Res} f=0 $. 

On the torus, it is more convenient to replace the plane singularity by the normalized second-kind differential \(P_2(z|\tau) = -\p_z^2 \log \theta_1(z|\tau)\), whose local behavior and periods are
\begin{equation}\label{P2defself}
    P_2(z|\tau)=\frac{1}{z^2}+\mathcal O(1),
    \qquad
    \oint_A P_2(z|\tau)dz=0,
    \qquad
    \oint_B P_2(z|\tau)dz=2\pi i.
\end{equation}
Thus, in correlation functions, we use
\begin{equation}\label{torusJJope}
    J^{(\lambda)}_1(z)J^{(\lambda)}_1(w)\sim -\kappa P_2(z-w|\tau),
    \qquad
    \bar J^{(\lambda)}_2(\bar z)\bar J^{(\lambda)}_2(\bar w)\sim
    -\bar\kappa \bar P_2(\bar z-\bar w|\bar\tau).
\end{equation}
For the anti-holomorphic kernel, the corresponding periods are
\begin{equation}\label{barP2periods}
    \oint_A \bar P_2(\bar z|\bar\tau)d\bar z=0,
    \qquad
    \oint_B \bar P_2(\bar z|\bar\tau)d\bar z=-2\pi i.
\end{equation}
Using this prescription and the Ward identity, the $J^{(\lambda)}_1(z)dz$ in the correlation function \eqref{JdzJJJ} should be the following meromorphic 1-form \cite{Eguchi:1986sb}:
\begin{equation}\label{J14.45}
\begin{aligned}
   \langle J_1(z)dz ~  J_1(z_1) \cd J_1(z_N) \rangle &= -\kappa \sum_{i=1}^N P_2(z-z_i|\tau) \langle J_1(z_1) \cd \widehat{J_1(z_i)} \cd J_1(z_N) \rangle \\
   & \qquad+ \langle 2\pi Q dz ~ J_1(z_1) \cd J_1(z_N) \rangle.
\end{aligned}
\end{equation}
In this equation we omit all the superscript $(\lambda)$ to avoid cluster of notations. Now, considering infinitesimal deformation, we still have
\begin{equation}
\begin{aligned}
    \p_\lambda Z_\lambda(\tau,\bar\tau;\chi,\bar\chi) &= - \Big\langle e^{\chi Q^{(\lambda)}+\bar \chi \bar Q^{(\lambda)}} \int_{\text{large-cycle}} d^2z ~J^{(\lambda)}_1(z)\bar J^{(\lambda)}_2(\bar z) \Big\rangle_{\lambda} \\
    &=- \frac{i}{2} \Big\langle e^{\chi Q^{(\lambda)}+\bar \chi \bar Q^{(\lambda)}} \left(
        \oint_AJ^{(\lambda)}_1\oint_B\bar J^{(\lambda)}_2
        -
        \oint_BJ^{(\lambda)}_1\oint_A\bar J^{(\lambda)}_2
    \right) \Big\rangle_{\lambda}.
\end{aligned}
\end{equation}
By expanding the chemical potential term, the key is to calculate the correlation function for some $J_1,~\bar J_2$ insertions:
\begin{equation}\label{4.54}
    \Big\langle J^{(\lambda)}_1(z_1)\cd J^{(\lambda)}_1(z_N) \bar J^{(\lambda)}_2(\bar w_1)\cd \bar J^{(\lambda)}_2(\bar w_M)  \left(
        \oint_AJ^{(\lambda)}_1\oint_B\bar J^{(\lambda)}_2
        -
        \oint_BJ^{(\lambda)}_1\oint_A\bar J^{(\lambda)}_2
    \right) \Big\rangle_\lambda.
\end{equation}
The integrals of $z_i$ and $\bar w_j$ are all along the $ A$-cycle, giving the charges $Q$ and $\bar Q$. Their integrals are regular among themselves, since
\begin{equation}
    \oint_A dz \oint_A dw ~P_2(z-w|\tau) = 0.
\end{equation}
This should be the case, because this is nothing but a $Q \circ Q = Q^2$ action on the cylinder. Thus, the only new input is the additional contribution in the $B$-cycle integrals
\begin{equation}
    Y^{(\lambda)} = \frac{1}{2\pi} \oint_B J^{(\lambda)}_1 ,\qquad \bar Y^{(\lambda)} = - \frac{1}{2\pi} \oint_B \bar J^{(\lambda)}_2.
\end{equation}
Because of \eqref{omega4.47} and \eqref{J14.45}, we have
\begin{equation}
\begin{aligned}
    Y^{(\lambda)}Q^{(\lambda)} &= \tau (Q^{(\lambda)})^2 -  \frac{1}{2\pi} i \kappa  \\
    \bar Y^{(\lambda)} \bar Q^{(\lambda)} &= \bar \tau (\bar Q^{(\lambda)})^2 + \frac{1}{2\pi} i \bar \kappa.  \\
\end{aligned}
\end{equation}
For more insertions, we can write
\begin{equation}\label{Ytorus4.52}
    \begin{aligned}
        Y^{(\lambda)} &= \tau Q^{(\lambda)} - \frac{i \kappa}{2\pi} \frac{\p}{\p Q^{(\lambda)}} \\
        \bar Y^{(\lambda)} &= \bar \tau \bar Q^{(\lambda)} + \frac{i \bar \kappa}{2\pi} \frac{\p}{\p \bar Q^{(\lambda)}}.
    \end{aligned}
\end{equation}
The large cycle integral part is
\begin{equation}
    -\frac{i}{2} \left(
        \oint_AJ^{(\lambda)}_1\oint_B\bar J^{(\lambda)}_2
        -
        \oint_BJ^{(\lambda)}_1\oint_A\bar J^{(\lambda)}_2
    \right) = \quad 2\pi^2 i (Q^{(\lambda)} \bar Y^{(\lambda)} - Y^{(\lambda)} \bar Q^{(\lambda)}).
\end{equation}
Combining all together, we have
\begin{equation}
\begin{aligned}
    &- \frac{i}{2} \Big\langle e^{\chi Q^{(\lambda)}+\bar \chi \bar Q^{(\lambda)}} \left(
        \oint_AJ^{(\lambda)}_1\oint_B\bar J^{(\lambda)}_2
        -
        \oint_BJ^{(\lambda)}_1\oint_A\bar J^{(\lambda)}_2
    \right) \Big\rangle_{\lambda} \\
    = & \Big\langle \z(4\pi^2 \tau_2 Q^{(\lambda)} \bar Q^{(\lambda)}  -\pi    \kappa \bar Q^{(\lambda)}  \frac{\p}{\p Q^{(\lambda)}}  -\pi \bar \kappa Q^{(\lambda)}  \frac{\p}{\p \bar Q^{(\lambda)}} \y) e^{\chi Q^{(\lambda)}+\bar \chi \bar Q^{(\lambda)}} \Big\rangle_\lambda \\
    = & \Big\langle \z( 4\pi^2 \tau_2 \p_\chi \p_{\bar \chi} -\pi \kappa \chi \p_{\bar \chi} - \pi \bar \kappa \bar \chi \p_\chi \y) e^{\chi Q^{(\lambda)} + \bar \chi \bar Q^{(\lambda)}} \Big\rangle_\lambda.
\end{aligned}
\end{equation}
Thus
\begin{equation}\label{3.55}
     \frac{\p}{\p \lambda}  Z_\lambda(\tau,\bar\tau;\chi,\bar\chi) = \z( 4\pi^2 \tau_2 \p_\chi \p_{\bar \chi} -\pi \kappa \chi \p_{\bar \chi} - \pi \bar \kappa \bar \chi \p_\chi \y)Z_\lambda(\tau,\bar\tau;\chi,\bar\chi).
\end{equation}

\subsubsection*{Higher genus}
The higher-genus generalization requires a normalized second-kind differential with a prescribed double pole and vanishing $A$-periods. We again fix a basis of holomorphic 1-forms as in \eqref{normalomega4.14}. For any $p \in \Sigma$, we choose a local coordinate $t=t_p$  on a neighbourhood $U_p$ of $p$ so that $t(p)=0$. Locally, the holomorphic 1-form $\omega_I$ can be written as
\begin{equation}
    \omega_I(z) = v_I (z;t)dt(z) ,\quad z\in U_p.
\end{equation}
The coefficient $v_I (z;t)$ is coordinate-dependent: let $t'$ be another local coordinate near $p$ and $t'(p)=0$, then locally
\begin{equation}
    t' = c t + \mathcal{O}(t^2),\qquad v_I(p;t') = \frac{1}{c} v_I(p;t).
\end{equation}
Now, consider a meromorphic 1-form $\eta$ with second order poles at $p_i,~i=1,\cd, N$. Locally near $p_i$, we choose a coordinate $t_i$ so that
\begin{equation}
    \eta = \z( \frac{a_i}{t_i^2} + \mathcal{O}(1) \y) dt_i.
\end{equation}
The coefficient $a_i$ above is also coordinate-dependent
\begin{equation}
    t_i' = c_i t_i + \mathcal{O}(t^2),\qquad a'_i = c_ia_i.
\end{equation}
Thus, the combination $a_i v_I(p_i;t_i)=:a_i v_I(p_i)$ is coordinate-independent. Consider the trivially vanishing integral
\begin{equation}
    0 = \int_{\Sigma_0 - \operatorname{Int} L} \eta \wedge \omega_I = \sum_{J=1}^g \z(  \oint_{A_J} \eta \oint_{B_J}\omega_I -  \oint_{B_J} \eta \oint_{A_J}\omega_I \y) - \oint_L f \omega_I, 
\end{equation}
where $f$ is the primitive of $\eta$ in $\Sigma_0 - \operatorname{Int} L$
\begin{equation}
    f = \int_P^z \eta.
\end{equation}
Then we have the higher genus generalization of \eqref{omega4.47}:
\begin{equation}\label{newomega4.62}
\begin{aligned}
        \oint_L f \omega_I  &= -2\pi i \sum_{i=1}^N a_i v_I(p_i)  \\
        &=  \Omega_{IJ} \oint_{A_J} \eta - \oint_{B_I} \eta.
\end{aligned}
\end{equation}
We also want to construct a generalization of $P_2(z|\tau)$. For this purpose, we note that for any $p$, there exists a meromorphic differential $\widetilde{\mathcal P}^{(t)}_p$ such that it only has a second order pole at $p$ with the local principal part
\begin{equation}\label{principalRHS}
    \widetilde{\mathcal P}^{(t)}_p = \frac{dt}{t^2} + \mathcal{O}(1),
\end{equation}
with a vanishing residue. This construction depends on the choice of the local coordinate $t$, and we add a superscript to indicate this. The $A$-periods are in general non-zero
\begin{equation}
    k_I = \oint_{A_I} \widetilde{\mathcal P}^{(t)}_{p}.
\end{equation}
Then, the following meromorphic 1-form has vanishing $A$-periods:
\begin{equation}
    \mathcal{P}^{(t)}_p = \widetilde{\mathcal{P}}^{(t)}_p - k_J \omega_J.
\end{equation}
Fixing the coordinate $t$, $\mathcal{P}^{(t)}_p $ is the unique meromorphic 1-form with the principal part in the RHS of \eqref{principalRHS} and vanishing $A$-periods: let $\mathcal P' $ be another example, then $\mathcal{P}^{(t)}_p - \mathcal P' $ is holomorphic
\begin{equation}
   \mathcal{P}^{(t)}_p - \mathcal P'  = \sum_J c_J \omega_J.
\end{equation}
Then
\begin{equation}
    0 = \oint_{A_I} (\mathcal{P}^{(t)}_p - \mathcal P') = c_I.
\end{equation}
From \eqref{newomega4.62}, the $B$-periods are 
\begin{equation}
    \oint_{B_I} \mathcal{P}^{(t)}_p = 2\pi i v_I(p;t).
\end{equation}
Now, we can write down the Ward identity that generalizes \eqref{J14.45}:
\begin{equation}
\begin{aligned}
   \langle J_1(z)dz ~  J_1(z_1) \cd J_1(z_N) \rangle &= -\kappa \sum_{i=1}^N \mathcal P^{(t_i)}_{z_i}(z) \langle J_1(z_1) \cd \widehat{J_1(z_i)} \cd J_1(z_N) \rangle \\
   & \qquad+ \langle 2\pi Q_I \omega_I(z) ~ J_1(z_1) \cd J_1(z_N) \rangle.
\end{aligned}
\end{equation}
We still define the charge operators $Q^{(\lambda)}_I$, $\bar{Q}^{(\lambda)}_I$, $Y^{(\lambda)}_I$, and $\bar Y^{(\lambda)}_I$ as cycle integrals of the currents. Then, the generalization of \eqref{Ytorus4.52} is clear
\begin{equation}
    \begin{aligned}
        Y^{(\lambda)}_I &= \Omega_{IJ} Q^{(\lambda)}_J- \frac{i\kappa}{2\pi} \frac{\p}{\p Q^{(\lambda)}_I} \\
        \bar Y^{(\lambda)}_I &= \bar \Omega_{IJ} \bar Q^{(\lambda)}_J +  \frac{i\bar \kappa}{2\pi} \frac{\p}{\p \bar Q^{(\lambda)}_I}. \\
    \end{aligned}
\end{equation}
By considering infinitesimal deformation of the partition function, we again encounter \eqref{4.54}. The large cycle integral part is
\begin{equation}
    -\frac{i}{2} \left(
        \oint_AJ^{(\lambda)}_1\oint_B\bar J^{(\lambda)}_2
        -
        \oint_BJ^{(\lambda)}_1\oint_A\bar J^{(\lambda)}_2
    \right) = \quad 2\pi^2 i \sum_{I=1}^g (Q_I^{(\lambda)} \bar Y_I^{(\lambda)} - Y_I^{(\lambda)} \bar Q_I^{(\lambda)}).
\end{equation}
Thus
\begin{equation}
     \frac{\p}{\p \lambda}  Z_\lambda = \z( 4\pi^2\p_\chi^T M \p_{\bar \chi} -\pi \kappa \chi^T \p_{\bar \chi} - \pi \bar \kappa \bar \chi^T \p_\chi \y)Z_\lambda.
\end{equation}

\subsubsection*{Modular covariance and the integration kernel}
Let us analyze the modular property of the differential operator of the $\lambda$-flow of partition functions. We begin with the torus case. It is obvious that the differential operator
\begin{equation}
    \mathcal L =  4\pi^2 \tau_2 \p_\chi \p_{\bar \chi} -\pi \kappa \chi \p_{\bar \chi} - \pi \bar \kappa \bar \chi \p_\chi 
\end{equation}
is not modular invariant. This should be the case, since the partition function with a non-vanishing chemical potential $Z_{0}(\tau,\bar \tau,\chi,\bar \chi)$ itself should not be modular invariant. For an affine $U(1)$ algebra, the partition function should transform as a Jacobi form
\begin{equation}
    Z_{0}(\tau',\bar \tau',\chi',\bar \chi') = \exp\z( - \frac{i\kappa c}{4\pi (c\tau +d)} \chi^2 + \frac{i\bar \kappa c}{4\pi (c\bar \tau +d)} \bar \chi^2 \y) Z_{0}(\tau,\bar \tau,\chi,\bar \chi),
\end{equation}
where the modular transformation is defined in \eqref{modulartorustransf}. Thus, modular covariance is preserved if
\begin{equation}
    \mathcal L' (Z_{0}(\tau',\bar \tau',\chi',\bar \chi'))= \exp\z( - \frac{i\kappa c}{4\pi (c\tau +d)} \chi^2 + \frac{i\bar \kappa c}{4\pi (c\bar \tau +d)} \bar \chi^2 \y) \mathcal L Z_{0}(\tau,\bar \tau,\chi,\bar \chi).
\end{equation}
This can be checked by direct computation. Moreover, one can introduce a non-holomorphic completion
\begin{equation}
    Z_{\operatorname{inv}} = F_s Z_{0} ,\qquad F_s = \exp\z( - \frac{1}{8\pi \tau_2} (\kappa \chi^2 + \bar \kappa \bar \chi^2 - 2s \chi \bar \chi) \y),
\end{equation}
where $s$ is a scheme choice since $\frac{\chi\bar\chi}{\tau_2}$ is already modular invariant. Then
\begin{equation}
    Z_{\operatorname{inv}}(\tau',\bar \tau', \chi',\bar \chi ) =  Z_{\operatorname{inv}}(\tau,\bar \tau,\chi,\bar \chi).
\end{equation}
We can also define a modular invariant differential operator acting on $Z_{\operatorname{inv}}$ as:
\begin{equation}
    \begin{aligned}
        \mathcal L^{(s)}_{\operatorname{inv}} &= F_s  \mathcal L F_s^{-1} \\
        \mathcal L^{(s)}_{\operatorname{inv}} &= 4\pi^2 \tau_2 \p_\chi \p_{\bar \chi} -\pi s(\chi\p_\chi + \bar \chi \p_{\bar \chi}) + \frac{s^2-\kappa\bar \kappa}{4\tau_2} \chi \bar \chi - \pi s. 
    \end{aligned}
\end{equation}
Now, we want to work out a kernel just as \eqref{kernel3.14}. It is convenient to consider the $s=0$ completion:
\begin{equation}
\begin{aligned}
    \widehat{Z}_\lambda &=  \exp\z( - \frac{1}{8\pi \tau_2} (\kappa \chi^2 + \bar \kappa \bar \chi^2 ) \y) Z_\lambda \\
    \p_\lambda \widehat{Z}_\lambda &= \z(4\pi^2 \tau_2 \p_\chi \p_{\bar \chi} - \frac{\kappa\bar \kappa}{4\tau_2} \chi \bar \chi \y) \widehat{Z}_\lambda.
\end{aligned}
\end{equation}
This differential equation can be integrated as
\begin{equation}
\begin{aligned}
    &\widehat{Z}_\lambda(\tau,\bar \tau,\chi,\bar \chi)  = \int_{\mathbb C} d^2\xi~ \widehat K_\lambda(\chi,\bar \chi;\xi,\bar \xi) \widehat{Z}_0(\tau,\bar \tau,\xi,\bar \xi) \\
    & \widehat K_\lambda(\chi,\bar \chi;\xi,\bar \xi)=\frac{\sqrt{\kappa\bar \kappa}}{4\pi^2\tau_2\sinh(\pi \sqrt{\kappa\bar \kappa} \lambda)}\exp \Bigg[ - \frac{\sqrt{\kappa\bar \kappa}\z((\chi\bar \chi + \xi \bar \xi) \cosh(\pi \sqrt{\kappa\bar \kappa} \lambda)-\chi \bar \xi - \bar \chi \xi \y)}{4\pi \tau_2 \sinh(\pi \sqrt{\kappa\bar \kappa} \lambda)} \Bigg]. 
\end{aligned}
\end{equation}
Then the kernel with no completion is
\begin{equation}
    \begin{aligned}
        {Z}_\lambda(\tau,\bar \tau,\chi,\bar \chi) & = \int_{\mathbb C} d^2\xi~ K_\lambda(\chi,\bar \chi;\xi,\bar \xi) {Z}_0(\tau,\bar \tau,\xi,\bar \xi)\\
        K_\lambda(\chi,\bar \chi;\xi,\bar \xi) &=  \exp\z(  \frac{\kappa (\chi^2-\xi^2) + \bar \kappa (\bar \chi^2-\bar \xi^2)  }{8\pi \tau_2} \y) \widehat K_\lambda(\chi,\bar \chi;\xi,\bar \xi).
    \end{aligned}
\end{equation}
For the higher genus case, introducing
\begin{equation}
    J = C\Omega + D,\qquad \bar J = C\bar \Omega + D.
\end{equation}
Then the Jacobi form transformed as
\begin{equation}
    Z_0(\Omega',\bar \Omega' ,\chi',\bar \chi') = \exp\z( - \frac{i\kappa}{4\pi} \chi J^{-1} C\chi^T + \frac{i\bar\kappa}{4\pi} \bar \chi \bar J^{-1} C\bar \chi^T \y).
\end{equation}
The non-holomorphic completion can be chosen as
\begin{equation}
    Z_{\operatorname{inv}} = F_s Z_0, \qquad F_s = \exp\z( - \frac{1}{8\pi} (\kappa \chi M^{-1} \chi^T + \bar \kappa \bar \chi M^{-1} \bar \chi^T - 2s \chi M^{-1} \bar \chi^T ) \y).
\end{equation}
We can still choose $s=0$ case to be $\widehat{Z}_{\operatorname{inv}}$, then the modular invariant differential operator is
\begin{equation}
    \mathcal L_{\operatorname{inv}} = 4\pi^2 \p_\chi^T M \p_{\bar \chi} - \frac{\kappa\bar \kappa}{4} \chi^T M^{-1} \bar \chi.
\end{equation}
Denote
\begin{equation}
    X = \begin{pmatrix}
        \chi \\
        \bar \chi
    \end{pmatrix} ,\qquad Y = \begin{pmatrix}
        \xi \\
        \bar \xi
    \end{pmatrix},\qquad \omega = \sqrt{\kappa\bar\kappa}, \qquad \theta=\pi \omega \lambda.
\end{equation}
Then the integral kernel is
\begin{equation}
    \begin{aligned}
        \widehat{Z}_\lambda (X) &= \int d^{2g} Y ~ \widehat K^{(g)}_\lambda(X;Y) \widehat Z_0(Y) \\
      \widehat  K^{(g)}_\lambda(X;Y) & = \frac{\z( \frac{\omega}{4\pi \sinh \theta} \y)^g}{\pi^g \det M} \exp\z[ - \frac{\omega}{4\pi} \z( \coth \theta (\chi^TM^{-1}\bar \chi + \xi^T M^{-1} \bar \xi ) - \frac{\chi^T M^{-1}\bar \xi + \xi^T M^{-1} \bar \chi}{\sinh \theta} \y) \y].
    \end{aligned}
\end{equation}
We also have the corresponding non-completed kernel
\begin{equation}
    K^{(g)}_\lambda(X;Y) = \big(F_0(X)\big)^{-1} \widehat K^{(g)}_\lambda(X;Y) F_0(Y).
\end{equation}

\subsection{Deformed spectrum}
The same flow equation also determines the deformation of the spectrum, provided that the relevant states do not mix under the deformation. The contribution of an individual state can be represented, via the state-operator correspondence, as a two-point function on the sphere with the Hamiltonian and charge insertions:
\begin{equation}
    \langle a | q^{L_0 - c/24} \bar q^{\bar L_0 - c/24} e^{\chi Q+\bar \chi \bar Q}  |a \rangle = \langle  V^{\dagger}_a(\infty) e^{-H(\tau,\bar \tau,;\chi,\bar \chi)} V_a(0) \rangle_{\mathbb CP^1}.
\end{equation}
As we turn on a $J\bar J$-deformation and turn it into a dress, the quantity above will become the corresponding contribution of $\ket{a^{(\lambda)}}$ to the partition function. More precisely, the contribution of $\ket{a^{(0)}}$ is
\begin{equation}
    W_a^{(0)}(\tau,\bar \tau;\chi,\bar \chi) = \exp \z[ 2\pi i\tau \z( h^{(0)} - \frac{c}{24} \y) - 2\pi i\bar \tau \z( \bar h^{(0)} - \frac{c}{24} \y) + \chi q^{(0)} + \bar \chi \bar q^{(0)} \y].
\end{equation}
Then, the corresponding contribution of $\ket{a^{(\lambda)}}$ is
\begin{equation}
    W_a^{(\lambda)}(\tau,\bar \tau;\chi,\bar \chi) 
    = \exp \z[ 2\pi i\tau \z( h^{(\lambda)} - \frac{c}{24} \y) - 2\pi i\bar \tau \z( \bar h^{(\lambda)} - \frac{c}{24} \y) + \chi q^{(\lambda)} + \bar \chi \bar q^{(\lambda)} \y].
\end{equation}
The deformed conformal weights and charges can thus be read by taking derivatives:
\begin{equation}
\begin{aligned}
   - \frac{1}{2\pi} \p_{\tau_2} \log W_a^{(\lambda)}(\tau,\bar \tau;\chi,\bar \chi) &= h^{(\lambda)} + \bar h^{(\lambda)} - \frac{c}{12} \\
    \frac{i}{2\pi} \p_{\tau_1} \log W_a^{(\lambda)}(\tau,\bar \tau;\chi,\bar \chi) &= h^{(\lambda)} - \bar h^{(\lambda)}  \\
    \p_{\chi}\log W_a^{(\lambda)}(\tau,\bar \tau;\chi,\bar \chi) &= q^{(\lambda)} \\
    \p_{\bar \chi}\log W_a^{(\lambda)}(\tau,\bar \tau;\chi,\bar \chi) &= \bar q^{(\lambda)}. \\
\end{aligned}
\end{equation}
Now, consider the differential equation
\begin{equation}
    \frac{\p}{\p \lambda} Z_\lambda = \sum_a \frac{\p}{\p \lambda} W_a^{(\lambda)} =\sum_a \z( 4\pi^2 \tau_2 \p_\chi \p_{\bar \chi} -\pi \kappa \chi \p_{\bar \chi} - \pi \bar \kappa \bar \chi \p_\chi \y)   W_a^{(\lambda)}.
\end{equation}
If $\{W^{(\lambda)}_a\}_a$ are linearly independent, namely there is no degeneracy in the $(L_0,\bar L_0, Q,\bar Q)$ spectrum, or the charge-running of the degenerate states is the same, then
\begin{equation}
    \p_\lambda W_a^{(\lambda)} =  \z( 4\pi^2 \tau_2 \p_\chi \p_{\bar \chi} -\pi \kappa \chi \p_{\bar \chi} - \pi \bar \kappa \bar \chi \p_\chi \y) W^{(\lambda)}_a.
\end{equation}
This implies
\begin{equation}\label{difffffffffqh}
\begin{aligned}
    \p_\lambda q^{(\lambda)} &= -\pi \kappa\bar q^{(\lambda)} \\
    \p_\lambda \bar q^{(\lambda)} &= -\pi \bar \kappa q^{(\lambda)} \\
    \p_\lambda h^{(\lambda)} &= -\pi q^{(\lambda)} \bar q^{(\lambda)} \\
    \p_\lambda \bar h^{(\lambda)} &= -\pi q^{(\lambda)} \bar q^{(\lambda)}. \\
\end{aligned}
\end{equation}
Denote
\begin{equation}
    \omega=\sqrt{\kappa \bar \kappa},\qquad \theta = \pi \omega \lambda.
\end{equation}
Then the solution is
\begin{equation}\label{qh4.9}
\begin{aligned}
    q^{(\lambda)} &= q^{(0)} \cosh \theta - \sqrt{\frac{\kappa}{\bar \kappa}}   \bar{q}^{(0)}\sinh \theta \\
    \bar q^{(\lambda)} &= \bar q^{(0)} \cosh \theta - \sqrt{\frac{\bar \kappa}{\kappa}}   {q}^{(0)}\sinh \theta \\
    h^{(\lambda)} &= h^{(0)} - \frac{q^{(0)}\bar q^{(0)}}{2\omega} \sinh(2\theta) + \frac{1}{4} \z( \frac{(q^{(0)})^2}{\kappa} + \frac{(\bar q^{(0)})^2}{\bar \kappa} \y)[\cosh(2\theta)-1] \\
    \bar h^{(\lambda)} &= \bar h^{(0)} - \frac{q^{(0)}\bar q^{(0)}}{2\omega} \sinh(2\theta) + \frac{1}{4} \z( \frac{(q^{(0)})^2}{\kappa} + \frac{(\bar q^{(0)})^2}{\bar \kappa} \y)[\cosh(2\theta)-1]. \\
\end{aligned}
\end{equation}
Note that the cases with $\kappa=0,\bar \kappa =0$ can be determined by taking limits. The result is
\begin{equation}\label{hq000000}
\begin{gathered}
    q^{(\lambda)} = q^{(0)},\qquad \bar q^{(\lambda)} = \bar q^{(0)} \\
    h^{(\lambda)} = h^{(0)} - \pi \lambda q^{(0)} \bar q^{(0)} ,\qquad \bar h^{(\lambda)} = \bar  h^{(0)} - \pi \lambda q^{(0)} \bar q^{(0)}.
\end{gathered}
\end{equation}

\subsection{$S$-transformation and dressed operators}
Consider the torus partition function with $\tau = i$ and vanishing chemical potential as a sum over states
\begin{equation}
    Z_\lambda = Z_\lambda(\tau =i,\bar \tau = -i;\chi = 0,\bar \chi=0)=\operatorname{Tr}_{\mathcal{H}_{\lambda,\loc}} e^{-H_\lambda}.
\end{equation}
Here, we add a subscript to emphasize that we only sum over local operators in the operator space. We have a kernel integral formula
\begin{equation}
    Z_\lambda = \int_{\mathbb C} d^2 \xi ~K_\lambda[\xi,\bar \xi] Z_0(\xi,\bar \xi).
\end{equation}
In terms of a partition sum, we have the $S$-transformation
\begin{equation}
       Z_0(\xi,\bar \xi) = \operatorname{Tr}_{\mathcal{H}_{0,\loc}} e^{-H_0 + \xi Q + \bar \xi \bar Q}  = \operatorname{Tr}_{\mathcal{H}_{0,(\xi,\bar\xi)}} e^{-H_0},
\end{equation}
where $\mathcal{H}_{0,(\xi,\bar\xi)}$ is the space of states on the cylinder with a line defect along the inverse time direction
\begin{equation}\label{Dxidefect}
    D_{\xi,\bar \xi} = \exp\z( \frac{\xi}{2\pi} \int_{-\mathbb R_t} J_1(z) dz - \frac{\bar \xi}{2\pi} \int_{-\mathbb R_t} \bar J_2(\bar z) d\bar z  \y).
\end{equation}
The kernel representation suggests the following interpretation:
\begin{equation}
\begin{aligned}
    Z_\lambda &= \operatorname{Tr}_{\mathcal{H}_{\lambda,\loc}} e^{-H_\lambda} = \int_{\mathbb C} d^2 \xi ~K_\lambda[\xi,\bar \xi] \operatorname{Tr}_{\mathcal{H}_{0,(\xi,\bar\xi)}} e^{-H_0} \\
   & = \operatorname{Tr}_{\Dr^\ren_{\lambda}( \mathcal{H}_{\lambda,\loc})} e^{-H_0}. 
\end{aligned}
\end{equation}
The last equal sign means that we dress each local operator in the $S_\lambda$-theory back to the $S_0$-theory; as in \eqref{rendress2.36}, the renormalized dressing maps the deformed stress tensor to the undeformed one, and so
\begin{equation}
    \Dr^\ren_\lambda(H_\lambda) = H_0.
\end{equation}
The dressing also preserves the correlation functions of local operators on the sphere, in particular
\begin{equation}
    \langle \psi_\lambda| e^{H_\lambda} |\psi_\lambda \rangle =\Dr_\lambda^\ren \z( \langle \psi_\lambda| e^{H_\lambda} |\psi_\lambda \rangle \y)   = \langle \Dr_\lambda^\ren( \psi_\lambda) | e^{H_0} | \Dr_\lambda^\ren(\psi_\lambda) \rangle. 
\end{equation}
Suppose we can build a linear map as a renormalized kernel integral of endpoints:
\begin{equation}\label{3.106}
\begin{aligned}
    \KE^\ren_\lambda : \mathcal{H}_{0} &\to  \mathcal{H}_{0} \\
                \mathcal{H}_{0,\loc} & \mapsto \Dr^\ren_{\lambda}( \mathcal{H}_{\lambda,\loc}) \\
                V_a^{(0)} & \mapsto
                \int_{\mathbb C} d^2 \xi ~ L_\lambda[a;\xi,\bar \xi] V_{a;(\xi,\bar \xi)}^{(0)} = \Dr^\ren_\lambda(V_a^{(\lambda)}).
\end{aligned}
\end{equation}
If this is possible, then we have on the sphere
\begin{equation}
    \langle X^{(\lambda)} \rangle_{\lambda,\mathbb C P^1} =\langle \Dr^\ren_\lambda( X^{(\lambda)} )\rangle_{0,\mathbb C P^1}  = \langle \KE^\ren_\lambda ( X^{(0)} )\rangle_{0,\mathbb C P^1}. 
\end{equation}
It turns out that the possible kernel $L_\lambda$ is simple if exists. The state-operator correspondence is
\begin{equation}
    \ket{a} = \lim_{z,\bar z \to 0} V_a(z,\bar z)  \ket{0}.
\end{equation}
Thus
\begin{equation}
\begin{aligned}
    \ket{ \Dr_\lambda^\ren (a^{(\lambda)})} &=  \int_{\mathbb C} d^2 \xi ~ L_\lambda[\xi,\bar \xi] ~\ket{a^{(0)};\xi,\bar \xi} \\
    \bra{ \Dr_\lambda^\ren (a^{(\lambda)})} &=  \int_{\mathbb C} d^2 \xi ~ L^*_\lambda[\xi,\bar \xi] ~\bra{a^{(0)};\xi,\bar \xi}.
\end{aligned}
\end{equation}
In the partition sum, the state $\ket{0}$ should have unit norm:
\begin{equation}\label{unitnorm4.22}
    \langle a | a \rangle = 1,\qquad \bra{a} L_0 \ket{a} = h_a.
\end{equation}
And we have
\begin{equation}\label{last4.23}
\begin{aligned}
   &\bra{\Dr_\lambda^\ren(a^{(\lambda)})} e^{-H_0} \ket{\Dr_\lambda^\ren(a^{(\lambda)})} \\
  =& \int d^2\xi_1 d^2\xi_2~ L^*_\lambda[a;\xi_1,\bar \xi_1] L_\lambda[a;\xi_2,\bar \xi_2] \langle a^{(0)};\xi_1,\bar \xi_1| e^{-H_0} | a^{(0)};\xi_2 ,\bar \xi_2 \rangle \\
  = & \int d^2\xi_1 d^2\xi_2~ L^*_\lambda[a;\xi_1,\bar \xi_1] L_\lambda[a;\xi_2,\bar \xi_2] \langle a^{(0)};\xi_1,\bar \xi_1| a^{(0)};\xi_2 ,\bar \xi_2 \rangle e^{-H_0(a^{(0)};\xi_2,\bar \xi_2)}. 
\end{aligned}
\end{equation}
Since the states in the partition sum are normalized as in \eqref{unitnorm4.22} rather than by a continuum delta function, a smooth kernel in defect space would not reproduce the state-by-state trace. This suggests that $L_\lambda[a;\xi,\bar \xi]$ must be itself a $\delta$-function in the $\xi$-space. That is, the dressed operator must take exactly the necessary $(\xi,\bar \xi)$-defect to form an endpoint operator that carries charges according to \eqref{qh4.9}. 

To be more explicit, let us take a closer look at the construction of the endpoint operator of the defect \eqref{Dxidefect}. Since the construction is fully defined in the undeformed operator space $\mathcal H_0$, we omit the superscript $(0)$ below. Let $V_{q,\bar q}$ be a charged operator with charges $(Q,\bar Q)= (q,\bar q)$ in the local Hilbert space $\mathcal{H}_{0,\loc}$; the endpoint operator $V_{(\xi,\bar \xi)}$ is required to have a non-trivial monodromy with charged local operators
\begin{equation}\label{monodromy4.80}
    V_{q,\bar q}(e^{2\pi i}z, e^{-2\pi i}\bar z) V_{(\xi,\bar \xi)} (0) = e^{\xi q + \bar \xi \bar q}  V_{q,\bar q}(z, \bar z) V_{(\xi,\bar \xi)} (0).
\end{equation}
We can introduce the endpoint operator
\begin{equation}
    T_{(\xi,\bar \xi)}(x) = \exp\z( \frac{\xi}{2\pi} \int_{\infty}^x J_1(z) dz - \frac{\bar \xi}{2\pi} \int_{\infty}^x \bar J_2(\bar z) d\bar z  \y).
\end{equation}
Then, other endpoint operators can be constructed by some normalized product of $T_{(\xi,\bar \xi)}$ and a local operator $V_{\loc} \in \mathcal{H}_{0,\loc}$. This fusion will change the $U(1)\times U(1)$ charges and the conformal weights of the undeformed operator $V_{\loc}$, depending on the parameter $(\xi,\bar \xi)$. Then, our criterion is that one should take the correct amount of $(\xi,\bar \xi)$ so that the operator $(T_{(\xi,\bar \xi)}V_{\loc})$ has the charges and conformal weights given in \eqref{qh4.9}. Then, this operator is the dressed operator that we are looking for, namely
\begin{equation}\label{3.114}
    (T_{(\xi,\bar \xi)}V^{(0)}_{\loc}) \simeq \Dr^\ren_\lambda (V_\loc^{(\lambda)}).
\end{equation}
For vertex operators taking the form of the exponential wave of the bosonization fields, this rule is true, since the dressing is just a linear recombination of the chiral and anti-chiral halves of the bosons, as we will see in \ref{app:rank-one-free-bosons}. Thus, determining the charges according to \eqref{qh4.9} will uniquely fix the vertex. However, if there are some derivatives in the vertex, which usually happens for descendant states, we should also construct the derivatives properly. We will also see examples of this in \eqref{4.91dress} and \eqref{Op4.96}.

\section{Examples}
We now test the formalism in explicit theories. The central task is to construct a renormalized dressing that satisfies \eqref{DrXXX2.35} and \eqref{rendress2.36}. It turns out that if the $U(1)\times U(1)$ algebra can be embedded into an $O(2,2)$ system, then the renormalized dressing can be chosen as the bare dressing of the $O(2,2)$ real bosons. As an application, we will also discuss the string theory in the TsT deformed background.

\subsection{$O(2,2)$ theories}
\label{app:rank-one-free-bosons}

For specific models, it is usually possible to bosonize the currents $J_1$ and $\bar J_2$. Thus, it is helpful to carefully study $O(2,2)$ theories. In this subsection, we will consider various examples containing two compact bosons
\begin{equation}
    X^i\sim X^i+2\pi,\qquad i=1,2.
\end{equation}
We will show that in this bosonization of $J_1$ and $\bar{J}_2$, the bare dressing of $X^i$ is the renormalized dressing. We will also verify the flow of the partition function in several different cases. 

\subsubsection*{General setups}

We will denote $\lambda$ as the flat coordinate of the deformation and $\mu$ as the non-flat coordinate which the action linearly depends on. The general action can be written as
\begin{equation}
    S_\mu
    =
    \frac1\pi\int d^2z~
    E_{\mu,ij}\p X^i\bp X^j,
    \qquad
    E_\mu=G_\mu+B_\mu,
\end{equation}
where
\begin{equation}
    G_\mu^T=G_\mu,\qquad B_\mu^T=-B_\mu .
\end{equation}
Locally,
\begin{equation}
    X^i(z,\bar z)=X_L^i(z)+X_R^i(\bar z),
\end{equation}
and the OPEs are
\begin{equation}
    \p X_L^i(z)\p X_L^j(w)
    \sim
    -\frac12\frac{(G_\mu^{-1})^{ij}}{(z-w)^2},
    \qquad
    \bp X_R^i(\bar z)\bp X_R^j(\bar w)
    \sim
    -\frac12\frac{(G_\mu^{-1})^{ij}}{(\bar z-\bar w)^2}.
\end{equation}
We choose two constant vectors \(u,v\), and define the deformation in the non-flat coordinate
\(\mu\) by
\begin{equation}
    E_\mu=E_0+\mu uv^T .
\end{equation}
Thus
\begin{equation}
    G_\mu
    =
    G_0+\frac{\mu}{2}\z(uv^T+vu^T\y),
    \qquad
    B_\mu
    =
    B_0+\frac{\mu}{2}\z(uv^T-vu^T\y).
\end{equation}
The unnormalized currents are
\begin{equation}
    j_u^{(\mu)}=u_i\p X_L^i,
    \qquad
    \bar j_v^{(\mu)}=v_i\bp X_R^i.
\end{equation}
We further define
\begin{equation}
    \mathcal A_\mu
    =
    \frac12u^TG_\mu^{-1}u,
    \qquad
    \mathcal B_\mu
    =
    \frac12v^TG_\mu^{-1}v,
    \qquad
    \mathcal C_\mu
    =
    \frac12u^TG_\mu^{-1}v .
\end{equation}
Then
\begin{equation}
    j_u^{(\mu)}(z)j_u^{(\mu)}(w)
    \sim
    -\frac{\mathcal A_\mu}{(z-w)^2},
\end{equation}
\begin{equation}
    \bar j_v^{(\mu)}(\bar z)\bar j_v^{(\mu)}(\bar w)
    \sim
    -\frac{\mathcal B_\mu}{(\bar z-\bar w)^2}.
\end{equation}
If we introduce also the fields \(v_iX_L^i\) and \(u_iX_R^i\), then
\begin{equation}
    u_i\p X_L^i(z)~v_j\p X_L^j(w)
    \sim
    -\frac{\mathcal C_\mu}{(z-w)^2},
\end{equation}
and
\begin{equation}
    v_i\bp X_R^i(\bar z)~u_j\bp X_R^j(\bar w)
    \sim
    -\frac{\mathcal C_\mu}{(\bar z-\bar w)^2}.
\end{equation}
The running of these coefficients can be solved explicitly in the non-flat coordinate $\mu$. It is useful to record the same coefficients in terms of their values at
\(\mu=0\). Define
\begin{equation}
    D_0=\mathcal A_0\mathcal B_0-\mathcal C_0^2,
\end{equation}
and
\begin{equation}
    \Delta_\mu
    :=
    \frac{\det G_\mu}{\det G_0}
    =
    1+2\mathcal C_0\mu-D_0\mu^2 .
\end{equation}
Then a direct inversion of \(G_\mu\) gives
\begin{equation}
    \mathcal A_\mu=\frac{\mathcal A_0}{\Delta_\mu},
    \qquad
    \mathcal B_\mu=\frac{\mathcal B_0}{\Delta_\mu},
    \qquad
    \mathcal C_\mu=\frac{\mathcal C_0-D_0\mu}{\Delta_\mu}.
\end{equation}
These formulae simply give the coefficients of the quadratic action at the point $\mu$ in terms of their values at $\mu=0$.

We now normalize the currents as
\begin{equation}
    J_1^{(\mu)}=a_\mu j_u^{(\mu)},
    \qquad
    \bar J_2^{(\mu)}=b_\mu\bar j_v^{(\mu)} .
\end{equation}
We choose the normalization so that
\begin{equation}
    a_\mu=a_0\sqrt{\Delta_\mu},
    \qquad
    b_\mu=b_0\sqrt{\Delta_\mu}.
\end{equation}
They satisfy
\begin{equation}
    \p_\mu\log a_\mu=\mathcal C_\mu,
    \qquad
    \p_\mu\log b_\mu=\mathcal C_\mu .
\end{equation}
With this normalization, the current algebra levels are independent of
\(\mu\):
\begin{equation}
    J_1^{(\mu)}(z)J_1^{(\mu)}(w)
    \sim
    -\frac{\kappa}{(z-w)^2},
    \qquad
    \bar J_2^{(\mu)}(\bar z)\bar J_2^{(\mu)}(\bar w)
    \sim
    -\frac{\bar\kappa}{(\bar z-\bar w)^2},
\end{equation}
where
\begin{equation}
    \kappa=a_0^2\mathcal A_0,
    \qquad
    \bar\kappa=b_0^2\mathcal B_0 .
\end{equation}
Denote normalized primitive fields as
\begin{equation}
    \Phi_1^{(\mu)}=a_\mu u_iX^i,
    \qquad
    \Phi_2^{(\mu)}=b_\mu v_iX^i .
\end{equation}
Introduce mixed OPE coefficients
\begin{equation}
    \p\Phi_1^{(\mu)}(z)\p\Phi_2^{(\mu)}(w)
    \sim
    -\frac{K_\mu}{(z-w)^2},
\end{equation}
and
\begin{equation}
    \bp\Phi_1^{(\mu)}(\bar z)\bp\Phi_2^{(\mu)}(\bar w)
    \sim
    -\frac{\bar K_\mu}{(\bar z-\bar w)^2}.
\end{equation}
In the present family of deformation, we have
\begin{equation}
    K_\mu=\bar K_\mu
    =
    a_\mu b_\mu\mathcal C_\mu
    =
    a_0b_0\left(\mathcal C_0-D_0\mu\right).
\end{equation}
The flat coordinate \(\lambda\) is defined by
\begin{equation}
    \frac{\p S_\lambda}{\p\lambda}
    =
    \int d^2z~J_1^{(\lambda)}\bar J^{(\lambda)}_2 .
\end{equation}
Since
\begin{equation}
    \frac{\p S_\mu}{\p\mu}
    =
    \frac1\pi\int d^2z~
    j_u^{(\mu)}\bar j_v^{(\mu)},
\end{equation}
we have
\begin{equation}
    \frac{d\mu}{d\lambda}
    =
    \pi a_\mu b_\mu
    =
    \pi a_0b_0\Delta_\mu .
\end{equation}
Now the flow of \(K_\mu\) follows immediately. From
\begin{equation}
    K_\mu=a_0b_0\left(\mathcal C_0-D_0\mu\right),
\end{equation}
we get
\begin{equation}
    \frac{dK_\mu}{d\mu}
    =
    -a_0b_0D_0 .
\end{equation}
On the other hand,
\begin{equation}
    K_\mu^2-\kappa\bar\kappa
    =
    a_0^2b_0^2
    \left[
        \left(\mathcal C_0-D_0\mu\right)^2
        -
        \mathcal A_0\mathcal B_0
    \right]
    =
    -a_0^2b_0^2D_0\Delta_\mu .
\end{equation}
Using \(d\mu/d\lambda=\pi a_0b_0\Delta_\mu\), we obtain a Riccati equation
\begin{equation}
    \frac{dK_\lambda}{d\lambda}
    =
    \pi\left(K_\lambda^2-\kappa\bar\kappa\right).
\end{equation}
The same equation holds for \(\bar K_\lambda\). Note that even if $\kappa=\bar \kappa=0$, there can be a non-trivial running of $K_\lambda$ as long as $K_0\neq 0$. The coordinate $\mu$ becomes flat if and only if $\kappa=\bar \kappa = K_0=\bar K_0 = 0$.

\subsubsection*{Renormalized dressing}
We now verify directly that the bare RBI dressing of the elementary fields \(X^i\) is already the renormalized dressing. We consider the relative dressing from \(S_{\mu+\delta\mu}\) to \(S_\mu\). From \eqref{dresssingle}, the first-order bare dressing of $X^i = X^i_L + X^i_R$ gives
\begin{equation}
    \Dr _{\mu+\delta\mu\to\mu}
    \left(X_L^{i,(\mu+\delta\mu)}\right)
    =
    X_L^{i,(\mu)}
    -
    \frac{\delta\mu}{2}
    (G_\mu^{-1}v)^i
    u_jX_L^{j,(\mu)}
    +
    \mathcal O(\delta\mu^2),
\end{equation}
and
\begin{equation}
    \Dr _{\mu+\delta\mu\to\mu}
    \left(X_R^{i,(\mu+\delta\mu)}\right)
    =
    X_R^{i,(\mu)}
    -
    \frac{\delta\mu}{2}
    (G_\mu^{-1}u)^i
    v_jX_R^{j,(\mu)}
    +
    \mathcal O(\delta\mu^2).
\end{equation}
Equivalently,
\begin{equation}
    X_L^{(\mu+\delta\mu)}
    \mapsto
    M_{L,\mu}X_L^{(\mu)},
    \qquad
    M_{L,\mu}
    =
    1-\frac{\delta\mu}{2}G_\mu^{-1}vu^T,
\end{equation}
and
\begin{equation}
    X_R^{(\mu+\delta\mu)}
    \mapsto
    M_{R,\mu}X_R^{(\mu)},
    \qquad
    M_{R,\mu}
    =
    1-\frac{\delta\mu}{2}G_\mu^{-1}uv^T.
\end{equation}
The first check is the OPE metric. Since
\begin{equation}
    G_{\mu+\delta\mu}
    =
    G_\mu
    +
    \frac{\delta\mu}{2}
    \left(uv^T+vu^T\right),
\end{equation}
we have
\begin{equation}
    G_{\mu+\delta\mu}^{-1}
    =
    G_\mu^{-1}
    -
    \frac{\delta\mu}{2}
    G_\mu^{-1}
    \left(uv^T+vu^T\right)
    G_\mu^{-1}
    +
    \mathcal O(\delta\mu^2).
\end{equation}
On the other hand,
\begin{equation}
\begin{aligned}
    M_{L,\mu}G_\mu^{-1}M_{L,\mu}^T
    &=
    G_\mu^{-1}
    -
    \frac{\delta\mu}{2}
    G_\mu^{-1}vu^TG_\mu^{-1}
    -
    \frac{\delta\mu}{2}
    G_\mu^{-1}uv^TG_\mu^{-1}
    +
    \mathcal O(\delta\mu^2)
    \\
    &=
    G_{\mu+\delta\mu}^{-1}
    +
    \mathcal O(\delta\mu^2).
\end{aligned}
\end{equation}
Similarly,
\begin{equation}
    M_{R,\mu}G_\mu^{-1}M_{R,\mu}^T
    =
    G_{\mu+\delta\mu}^{-1}
    +
    \mathcal O(\delta\mu^2).
\end{equation}
Therefore the dressed elementary fields have exactly the OPE metric of the \(S_{\mu+\delta\mu}\)-theory. The second check is the normalized current. From
\begin{equation}
    M_{L,\mu}^Tu
    =
    \left(1-\mathcal C_\mu\delta\mu\right)u
    +
    \mathcal O(\delta\mu^2),
\end{equation}
we find
\begin{equation}
\begin{aligned}
    \Dr _{\mu+\delta\mu\to\mu}
    \left(J_1^{(\mu+\delta\mu)}\right)
    &=
    a_{\mu+\delta\mu}
    u_i\p
    \Dr _{\mu+\delta\mu\to\mu}
    \left(X_L^{i,(\mu+\delta\mu)}\right)
    \\
    &=
    a_{\mu+\delta\mu}
    \left(1-\mathcal C_\mu\delta\mu\right)
    u_i\p X_L^{i,(\mu)}
    +
    \mathcal O(\delta\mu^2).
\end{aligned}
\end{equation}
Using
\begin{equation}
    a_{\mu+\delta\mu}
    =
    a_\mu
    \left(1+\mathcal C_\mu\delta\mu\right)
    +
    \mathcal O(\delta\mu^2),
\end{equation}
we get
\begin{equation}
    \Dr _{\mu+\delta\mu\to\mu}
    \left(J_1^{(\mu+\delta\mu)}\right)
    =
    J_1^{(\mu)}
    +
    \mathcal O(\delta\mu^2).
\end{equation}
The right-moving calculation is identical:
\begin{equation}
    \Dr _{\mu+\delta\mu\to\mu}
    \left(\bar J_2^{(\mu+\delta\mu)}\right)
    =
    \bar J_2^{(\mu)}
    +
    \mathcal O(\delta\mu^2).
\end{equation}
Thus the bare dressing has the two defining properties expected from the
renormalized dressing: it maps the OPE metric of the elementary fields to the
OPE metric of the deformed theory, and it maps the normalized currents in
\(S_{\mu+\delta\mu}\) back to the normalized currents in \(S_\mu\). Hence, in
the elementary \(X^i\)-basis,
\begin{equation}
    \Dr _{\mu+\delta\mu\to\mu}
    \left(X^i\right)
    =
    \Dr^{\ren}_{\mu+\delta\mu\to\mu}
    \left(X^i\right)
    +
    \mathcal O(\delta\mu^2).
\end{equation}
Understanding this, we can integrate to get the finite renormalized dressing
\begin{equation}
    \Dr_\mu^\ren(X^{(\mu)}_L) = M_L(\mu) X^{(0)}_L,\qquad \Dr_\mu^\ren(X^{(\mu)}_R) = M_R(\mu) X^{(0)}_R,
\end{equation}
where
\begin{equation}
\begin{aligned}
& M_L(\mu)=\mathbf{1}-\frac{\mu}{2 \sqrt{\Delta_\mu}}\left(G_0^{-1} v-\frac{\mu \mathcal{B}_0}{1+\mathcal{C}_0 \mu+\sqrt{\Delta_\mu}} G_0^{-1} u\right) u^T, \\
& M_R(\mu)=\mathbf{1}-\frac{\mu}{2 \sqrt{\Delta_\mu}}\left(G_0^{-1} u-\frac{\mu \mathcal{A}_0}{1+\mathcal{C}_0 \mu+\sqrt{\Delta_\mu}} G_0^{-1} v\right) v^T .
\end{aligned}
\end{equation}

\subsubsection*{Spectrum and partition function}
Now, let us discuss partition functions. For compact bosons, the local vertex operators are labeled by
\begin{equation}
    n_i\in\mathbb Z,
    \qquad
    w^i\in\mathbb Z.
\end{equation}
We define
\begin{equation}
    P_L^{(\mu)}=n+E_\mu w,
    \qquad
    P_R^{(\mu)}=n-E_\mu^Tw.
\end{equation}
The zero-mode weights are
\begin{equation}
    h_{0,n,w}^{(\mu)}
    =
    \frac14
    (P_L^{(\mu)})^TG_\mu^{-1}P_L^{(\mu)},
    \qquad
    \bar h_{0,n,w}^{(\mu)}
    =
    \frac14
    (P_R^{(\mu)})^TG_\mu^{-1}P_R^{(\mu)}.
\end{equation}
Including oscillators,
\begin{equation}
    h_{n,w}^{(\mu)}=h_{0,n,w}^{(\mu)}+N,
    \qquad
    \bar h_{n,w}^{(\mu)}=\bar h_{0,n,w}^{(\mu)}+\bar N .
\end{equation}
The spin is independent of \(\mu\):
\begin{equation}
    h_{n,w}^{(\mu)}-\bar h_{n,w}^{(\mu)}
    =
    n^Tw+N-\bar N.
\end{equation}
The normalized charges are
\begin{equation}
    Q_{n,w}^{(\mu)}
    =
    \frac{a_\mu}{2}
    u^TG_\mu^{-1}P_L^{(\mu)},
\end{equation}
and
\begin{equation}
    \bar Q_{n,w}^{(\mu)}
    =
    \frac{b_\mu}{2}
    v^TG_\mu^{-1}P_R^{(\mu)}.
\end{equation}
Direct differentiation gives the state-by-state flow
\begin{equation}
    \frac{dQ}{d\lambda}
    =
    -\pi\kappa\bar Q,
    \qquad
    \frac{d\bar Q}{d\lambda}
    =
    -\pi\bar\kappa Q,
\end{equation}
and
\begin{equation}
    \frac{d}{d\lambda}(h+\bar h)
    =
    -2\pi Q\bar Q,
    \qquad
    \frac{d}{d\lambda}(h-\bar h)=0.
\end{equation}
This is consistent with \eqref{difffffffffqh}. The torus partition function with chemical potentials is
\begin{equation}
    Z_\lambda(\tau,\bar\tau;\chi,\bar\chi)
    =
    \frac1{|\eta(\tau)|^4}
    \sum_{n,w}
    e^{\chi Q_{n,w}^{(\lambda)}
    +
    \bar\chi\bar Q_{n,w}^{(\lambda)}}
    q^{h_{0,n,w}^{(\lambda)}}
    \bar q^{\bar h_{0,n,w}^{(\lambda)}} .
\end{equation}
Therefore each summand satisfies
\begin{equation}
    \p_\lambda
    \z(
        e^{\chi Q+\bar\chi\bar Q}
        q^{h_0}
        \bar q^{\bar h_0}
    \y)
    =
    \z(
        4\pi^2\tau_2 Q\bar Q
        -
        \pi\kappa\chi\bar Q
        -
        \pi\bar\kappa\bar\chi Q
    \y)
    e^{\chi Q+\bar\chi\bar Q}
    q^{h_0}
    \bar q^{\bar h_0}.
\end{equation}
After summing over states, we obtain \eqref{3.55}: 
\begin{equation}
    \p_\lambda Z_\lambda
    =
    \z(
        4\pi^2\tau_2\p_\chi\p_{\bar\chi}
        -
        \pi\kappa\chi\p_{\bar\chi}
        -
        \pi\bar\kappa\bar\chi\p_\chi
    \y)
    Z_\lambda. 
\end{equation}
Let us now check vertex operators. The local vertex operators in the
\(S_{\mu+\delta\mu}\)-theory are
\begin{equation}
    V_{n,w}^{(\mu+\delta\mu)}
    =
    :
    \exp\left[
        i(P_L^{(\mu+\delta\mu)})^TX_L^{(\mu+\delta\mu)}
        +
        i(P_R^{(\mu+\delta\mu)})^TX_R^{(\mu+\delta\mu)}
    \right]
    :_{\mu+\delta\mu},
\end{equation}
where
\begin{equation}
    P_L^{(\mu+\delta\mu)}
    =
    n+E_{\mu+\delta\mu}w,
    \qquad
    P_R^{(\mu+\delta\mu)}
    =
    n-E_{\mu+\delta\mu}^Tw.
\end{equation}
Using the \(X\)-basis dressing, we obtain
\begin{equation}
\begin{aligned}
    \Dr^{\ren}_{\mu+\delta\mu\to\mu}
    \left(
        V_{n,w}^{(\mu+\delta\mu)}
    \right)
    &=
    :
    \exp\left[
        i(P_L^{(\mu+\delta\mu)})^T
        M_{L,\mu}X_L^{(\mu)}
    \right.
    \\
    &\hspace{2.7cm}
    \left.
        +
        i(P_R^{(\mu+\delta\mu)})^T
        M_{R,\mu}X_R^{(\mu)}
    \right]
    :_{\mu}
    \\
    &\qquad\times
    \left(\text{field-strength factor}\right)
    +
    \mathcal O(\delta\mu^2).
\end{aligned}
\end{equation}
The field-strength factor comes from changing the normal-ordering
prescription. It is proportional to the operator itself and does not affect
its conformal weights or \(U(1)\times U(1)\) charges.

Equivalently, the dressed operator in the \(S_\mu\)-theory has left and
right momenta
\begin{equation}
    M_{L,\mu}^TP_L^{(\mu+\delta\mu)},
    \qquad
    M_{R,\mu}^TP_R^{(\mu+\delta\mu)}.
\end{equation}
Expanding these momenta to first order gives a useful local form. Since
\begin{equation}
    E_{\mu+\delta\mu}
    =
    E_\mu+\delta\mu\,uv^T,
\end{equation}
we have
\begin{equation}
    P_L^{(\mu+\delta\mu)}
    =
    P_L^{(\mu)}
    +
    \delta\mu\,u(v^Tw),
    \qquad
    P_R^{(\mu+\delta\mu)}
    =
    P_R^{(\mu)}
    -
    \delta\mu\,v(u^Tw).
\end{equation}
Using
\begin{equation}
    P_L^{(\mu)}-P_R^{(\mu)}
    =
    2G_\mu w,
\end{equation}
one finds
\begin{equation}
\begin{aligned}
    M_{L,\mu}^TP_L^{(\mu+\delta\mu)}
    &=
    P_L^{(\mu)}
    -
    \frac{\delta\mu}{2}
    u\,v^TG_\mu^{-1}P_R^{(\mu)}
    +
    \mathcal O(\delta\mu^2),
\end{aligned}
\end{equation}
and
\begin{equation}
\begin{aligned}
    M_{R,\mu}^TP_R^{(\mu+\delta\mu)}
    &=
    P_R^{(\mu)}
    -
    \frac{\delta\mu}{2}
    v\,u^TG_\mu^{-1}P_L^{(\mu)}
    +
    \mathcal O(\delta\mu^2).
\end{aligned}
\end{equation}
Thus the dressed vertex is, up to the field-strength factor,
\begin{equation}
\begin{aligned}
    \Dr^{\ren}_{\mu+\delta\mu\to\mu}
    \left(
        V_{n,w}^{(\mu+\delta\mu)}
    \right)
    &=
    :
    \exp\left[
        i\left(
            P_L^{(\mu)}
            -
            \frac{\delta\mu}{2}
            u\,v^TG_\mu^{-1}P_R^{(\mu)}
        \right)^TX_L^{(\mu)}
    \right.
    \\
    &\hspace{2.7cm}
    \left.
        +
        i\left(
            P_R^{(\mu)}
            -
            \frac{\delta\mu}{2}
            v\,u^TG_\mu^{-1}P_L^{(\mu)}
        \right)^TX_R^{(\mu)}
    \right]
    :_{\mu}
    \\
    &\qquad\qquad
    +
    \mathcal O(\delta\mu^2).
\end{aligned}
\end{equation}
This expression should not be interpreted as an ordinary lattice vertex of
the \(S_\mu\)-theory with integral labels \((n,w)\). Rather, it is the
dressed operator obtained by transporting the \(S_{\mu+\delta\mu}\)-vertex
back to the \(S_\mu\)-theory.

We now compute its conformal weights in the \(S_\mu\)-theory. The holomorphic
weight of the dressed vertex is
\begin{equation}
\begin{aligned}
    h_\mu
    \left[
        \Dr^{\ren}_{\mu+\delta\mu\to\mu}
        \left(
            V_{n,w}^{(\mu+\delta\mu)}
        \right)
    \right]
    &=
    \frac14
    (P_L^{(\mu+\delta\mu)})^T
    M_{L,\mu}G_\mu^{-1}M_{L,\mu}^T
    P_L^{(\mu+\delta\mu)}
    \\
    &=
    \frac14
    (P_L^{(\mu+\delta\mu)})^T
    G_{\mu+\delta\mu}^{-1}
    P_L^{(\mu+\delta\mu)}
    +
    \mathcal O(\delta\mu^2)
    \\
    &=
    h_{n,w}^{(\mu+\delta\mu)}
    +
    \mathcal O(\delta\mu^2).
\end{aligned}
\end{equation}
Similarly,
\begin{equation}
\begin{aligned}
    \bar h_\mu
    \left[
        \Dr^{\ren}_{\mu+\delta\mu\to\mu}
        \left(
            V_{n,w}^{(\mu+\delta\mu)}
        \right)
    \right]
    &=
    \frac14
    (P_R^{(\mu+\delta\mu)})^T
    M_{R,\mu}G_\mu^{-1}M_{R,\mu}^T
    P_R^{(\mu+\delta\mu)}
    \\
    &=
    \frac14
    (P_R^{(\mu+\delta\mu)})^T
    G_{\mu+\delta\mu}^{-1}
    P_R^{(\mu+\delta\mu)}
    +
    \mathcal O(\delta\mu^2)
    \\
    &=
    \bar h_{n,w}^{(\mu+\delta\mu)}
    +
    \mathcal O(\delta\mu^2).
\end{aligned}
\end{equation}
Thus the dressed vertex in the \(S_\mu\)-theory has exactly the conformal
weights of the original vertex in the \(S_{\mu+\delta\mu}\)-theory.

The same check works for the normalized charges. In the \(S_\mu\)-theory, the
charge of the dressed vertex is
\begin{equation}
\begin{aligned}
    Q_\mu
    \left[
        \Dr^{\ren}_{\mu+\delta\mu\to\mu}
        \left(
            V_{n,w}^{(\mu+\delta\mu)}
        \right)
    \right]
    &=
    \frac{a_\mu}{2}
    u^TG_\mu^{-1}M_{L,\mu}^TP_L^{(\mu+\delta\mu)}.
\end{aligned}
\end{equation}
From the current check above, we have
\begin{equation}
    a_{\mu+\delta\mu}M_{L,\mu}^Tu
    =
    a_\mu u
    +
    \mathcal O(\delta\mu^2),
\end{equation}
or equivalently
\begin{equation}
    a_{\mu+\delta\mu}u^TM_{L,\mu}
    =
    a_\mu u^T
    +
    \mathcal O(\delta\mu^2).
\end{equation}
Therefore
\begin{equation}
\begin{aligned}
    Q_\mu
    \left[
        \Dr^{\ren}_{\mu+\delta\mu\to\mu}
        \left(
            V_{n,w}^{(\mu+\delta\mu)}
        \right)
    \right]
    &=
    \frac{a_{\mu+\delta\mu}}{2}
    u^TM_{L,\mu}G_\mu^{-1}M_{L,\mu}^T
    P_L^{(\mu+\delta\mu)}
    +
    \mathcal O(\delta\mu^2)
    \\
    &=
    \frac{a_{\mu+\delta\mu}}{2}
    u^TG_{\mu+\delta\mu}^{-1}
    P_L^{(\mu+\delta\mu)}
    +
    \mathcal O(\delta\mu^2)
    \\
    &=
    Q_{n,w}^{(\mu+\delta\mu)}
    +
    \mathcal O(\delta\mu^2).
\end{aligned}
\end{equation}
Similarly, using
\begin{equation}
    b_{\mu+\delta\mu}M_{R,\mu}^Tv
    =
    b_\mu v
    +
    \mathcal O(\delta\mu^2),
\end{equation}
we obtain
\begin{equation}
\begin{aligned}
    \bar Q_\mu
    \left[
        \Dr^{\ren}_{\mu+\delta\mu\to\mu}
        \left(
            V_{n,w}^{(\mu+\delta\mu)}
        \right)
    \right]
    &=
    \frac{b_\mu}{2}
    v^TG_\mu^{-1}M_{R,\mu}^TP_R^{(\mu+\delta\mu)}
    \\
    &=
    \frac{b_{\mu+\delta\mu}}{2}
    v^TG_{\mu+\delta\mu}^{-1}
    P_R^{(\mu+\delta\mu)}
    +
    \mathcal O(\delta\mu^2)
    \\
    &=
    \bar Q_{n,w}^{(\mu+\delta\mu)}
    +
    \mathcal O(\delta\mu^2).
\end{aligned}
\end{equation}

Therefore the vertex dressing induced by the \(X\)-basis bare dressing has
the defining property of the renormalized dressing:
\begin{equation}
\begin{gathered}
    h_\mu\left[
        \Dr^{\ren}_{\mu+\delta\mu\to\mu}
        \left(V_{n,w}^{(\mu+\delta\mu)}\right)
    \right]
    =
    h_{n,w}^{(\mu+\delta\mu)}
    +
    \mathcal O(\delta\mu^2),
    \\
    \bar h_\mu\left[
        \Dr^{\ren}_{\mu+\delta\mu\to\mu}
        \left(V_{n,w}^{(\mu+\delta\mu)}\right)
    \right]
    =
    \bar h_{n,w}^{(\mu+\delta\mu)}
    +
    \mathcal O(\delta\mu^2),
    \\
    Q_\mu\left[
        \Dr^{\ren}_{\mu+\delta\mu\to\mu}
        \left(V_{n,w}^{(\mu+\delta\mu)}\right)
    \right]
    =
    Q_{n,w}^{(\mu+\delta\mu)}
    +
    \mathcal O(\delta\mu^2),
    \\
    \bar Q_\mu\left[
        \Dr^{\ren}_{\mu+\delta\mu\to\mu}
        \left(V_{n,w}^{(\mu+\delta\mu)}\right)
    \right]
    =
    \bar Q_{n,w}^{(\mu+\delta\mu)}
    +
    \mathcal O(\delta\mu^2).
\end{gathered}
\end{equation}
Hence, the dressing of vertex operators is not an additional assumption. It is the dressing induced by the bare dressing of the elementary fields \(X^i(z,\bar z)\), together with the same field-strength renormalization that removes the diagonal double contraction.

\subsubsection*{Flat coordinate in no-self-interaction cases}\label{lcexample1}
Now, let us consider the cases with $\kappa=\bar \kappa=0$. We cannot use $\kappa$ and $\bar \kappa$ to normalize the OPE coefficients of the currents $J^{(\lambda)}_1$ and $\bar J^{(\lambda)}_2$. However, the renormalized dressing is defined with reference to the local primitive fields $\Phi^{(\lambda)}_1$ and $\Phi^{(\lambda)}_2$. Thus, although $\mathcal A_\lambda$ and $\mathcal B_\lambda$ vanish, there is still a running controlled by $\mathcal C_\lambda$ or $K_\lambda$ defined in the previous discussion. Let us use two simple examples to show this mechanism. 

First, consider
\begin{equation}
    S_\mu
    =
    \frac1\pi\int d^2z~
    \z[
        (t+\mu)\p\varphi_+\bp\varphi_-
        +
        t\p\varphi_-\bp\varphi_+
    \y].
\end{equation}
Thus, in the basis \((+,-)\),
\begin{equation}
    E_\mu
    =
    \begin{pmatrix}
        0&t+\mu\\
        t&0
    \end{pmatrix}.
\end{equation}
The metric is
\begin{equation}
    G_\mu
    =
    \begin{pmatrix}
        0& \frac{1}{2\mathcal C_\mu}\\
        \frac{1}{2\mathcal C_\mu}&0
    \end{pmatrix},
    \qquad
    \mathcal{C}_\mu=2t+\mu.
\end{equation}
Moreover, we have
\begin{equation}
    a_\mu = b_\mu = 1+\frac{\mu}{2t},\qquad K_\mu = \frac{1}{2t} + \frac{\mu}{4t^2}.
\end{equation}
Thus, the flat coordinate should be
\begin{equation}
    \frac{d\mu}{d\lambda} = \pi a_\mu b_\mu \quad \Rightarrow \quad \lambda = \frac{\mu}{\pi(1+\frac{\mu}{2t})}.
\end{equation}
The normalized currents should be non-trivially dependent on $\mu$:
\begin{equation}
    J^{(\mu)}
    =
    \z(1+\frac{\mu}{2t} \y) \p\varphi_{+,L},
    \qquad
    \bar J^{(\mu)}
    =
    \z(1+\frac{\mu}{2t} \y)  \bp\varphi_{-,R}.
\end{equation}
Moreover, as we have calculated in the general case, only using the flat coordinate $\lambda$ could one obtain \eqref{flowZ3.12}. 

Finally, let us consider
\begin{equation}\label{action4.86}
    S_\mu
    =
    \frac1\pi\int d^2z~
    \z[
        \mu\p\varphi_+\bp\varphi_+
        +
        t\p\varphi_+\bp\varphi_-
        +
        t\p\varphi_-\bp\varphi_+
    \y].
\end{equation}
In this example, $\mu$ is a genuine flat coordinate
\begin{equation}
    \frac{d\mu}{d\lambda}=\pi,
    \qquad
    \mu=\pi\lambda .
\end{equation}
And the dressing becomes linear
\begin{equation}
\begin{aligned}
    \Dr_{\nu \to \mu}^\ren (\varphi_-^{(\nu)}) &= \varphi_-^{(\mu)} - \frac{\nu-\mu}{2t} \varphi_+^{(\mu)} \\
    \Dr_{\nu \to \mu}^\ren (\varphi_+^{(\nu)}) &=\varphi_+^{(\mu)}.
\end{aligned}
\end{equation}
This example also exhibits an important subtlety for derivative primaries. In the undeformed theory, consider the accidental primary
\begin{equation}
    \mathcal{O}^{(0)}_{n,p} = \z( \p \varphi^{(0)}_- \y)^n e^{i2tp \varphi^{(0)}_-},\qquad h^{(0)} = n,\qquad q^{(0)}= \bar{q}^{(0)} = p.
\end{equation}
After the deformation, the charges should be
\begin{equation}
    q^{(\mu)} = q^{(0)} , \qquad h^{(\mu)} = h^{(0)} - \mu q^{(0)}\bar q^{(0)} = n-\mu p^2.
\end{equation}
The dressed operator should be of the form\footnote{The derivatives are normally ordered with each other and the exponential vertex. }
\begin{equation}\label{4.91dress}
    \Dr^\ren_\mu(\mathcal{O}^{(\mu)}_{n,p} ) = \z( \p \varphi^{(0)}_- + \cd  \y)^n e^{i(2tp \varphi^{(0)}_- - \mu p \varphi_+^{(0)} )},\qquad h^{(\mu)}=n-\mu p^2.
\end{equation}
In addition to shifting the exponent, the derivative term also needs to be adjusted so that it remains a primary field. For instance, for $n=1$
\begin{equation}
    \Dr^\ren_\mu(\mathcal{O}^{(\mu)}_{1,p} ) = \z( \p \varphi^{(0)}_- + \frac{\mu}{2t} \p \varphi_+^{(0)}  \y) e^{i(2tp \varphi^{(0)}_- - \mu p \varphi_+^{(0)} )}.
\end{equation}
One can check that this is indeed a primary in $S_0$-theory. Dressing the right-hand side inversely with $\Dr_{0\to\mu}^{\ren}$ gives a primary in the deformed theory:\begin{equation}
    \mathcal{O}^{(\mu)}_{1,p} = \z( \p \varphi^{(\mu)}_- + \frac{\mu}{t} \p \varphi_+^{(\mu)}  \y) e^{i2tp \varphi^{(\mu)}_- }.
\end{equation}
The construction for $n=2$ is not trivial, because the naive generalization is not a primary:
\begin{equation}
     \z( \p \varphi^{(\mu)}_- + \frac{\mu}{t} \p \varphi_+^{(\mu)}  \y)^2 e^{i2tp \varphi^{(\mu)}_- }.
\end{equation}
It lives in the level-2 descendants of the exponential vertex, thus one way to cure the non-primariness is to add other level-2 descendants so that both $L_1,L_2$ kill the combination. That is, we consider adding
\begin{equation}
    L_{-2}V_p, \qquad (L_{-1})^2 V_p = \p^2 V_p,\qquad V_p = e^{i2tp \varphi^{(\mu)}_- }.
\end{equation}
We find the following unique solution
\begin{equation}\label{Op4.96}
    \begin{gathered}
        \mathcal{O}^{(\mu)}_{2,p} = \z(\p \varphi^{(\mu)}_- + \frac{\mu}{t} \p \varphi_+^{(\mu)}\y)^2 V_p + \frac{\mu}{2 t^2 B} \z[ L_{-2} V_p - \frac{3}{2(1-2\mu p^2)} L_{-1}^2 V_p \y] \\
         B = \frac{1+3\mu p^2 + 8\mu^2 p^4}{1-2\mu p^2}. 
    \end{gathered}
\end{equation}
One can also dress this operator to obtain a primary in $S_0$-theory using the renormalized dressing:
\begin{equation}
\begin{aligned}
    \Dr^\ren_\mu( e^{i2tp \varphi^{(\mu)}_- } ) &= e^{i2tp \varphi^{(0)}_- - \mu p \varphi_+^{(0)} }  \\
    \Dr^\ren_\mu\z(\p \varphi^{(\mu)}_- + \frac{\mu}{t} \p \varphi_+^{(\mu)} \y) &=  \p \varphi^{(0)}_- + \frac{\mu}{2t} \p \varphi_+^{(0)}. 
\end{aligned}
\end{equation}
Thus, to build a map $\KE^\ren_\lambda$ as in \eqref{3.106} from primaries carrying derivatives to the corresponding dressed operator that is still a primary is highly non-trivial, although the dressing map $\Dr^\ren_\lambda$ is the same as $X^i$ and exponential fields.

\subsection{Strings on TsT background}
\label{subsec:TsT-real-O22-corrected}

We follow the AdS$_3$ convention and the Wakimoto construction in \cite{Knighton:2023mhq}. The undeformed theory is
\begin{equation}
S=\frac{1}{2 \pi} \int_{\Sigma} \mathrm{d}^2 z\left(\frac{1}{2} \partial \Phi \bar{\partial} \Phi +\beta \bar{\partial} \gamma+\bar{\beta} \partial \bar{\gamma}-\frac{1}{k} \beta \bar{\beta} e^{-Q \Phi }-\frac{Q}{4} R \Phi \right),
\end{equation}
where the background charge is $Q = \sqrt{2/(k-2)}$, and $k$ is the level of the $\mathrm{SL}(2,\mathbb R)$ WZW model. In the large $\Phi$ limit, the worldsheet theory becomes free, and the OPEs of the Wakimoto variables become
\begin{equation}
    \beta(z)\gamma(w)\sim-\frac1{z-w},
    \qquad
    \Phi(z,\bar z)\Phi(w,\bar w)\sim-\log|z-w|^2 .
\end{equation}
The stress tensor is
\begin{equation}
    T
    =
    -\frac12(\p\Phi_L)^2
    -
    \frac{Q}{2}\p^2\Phi_L
    -
    \beta\p\gamma. 
\end{equation}
It is convenient to further bosonize the $\beta\gamma$-system
\begin{equation}
    \beta = e^{\varphi_L + \theta_L},\qquad \gamma = e^{-\varphi_L - \theta_L},
\end{equation}
where
\begin{equation}
    \varphi_L(z) \varphi_L(w) \sim -\log (z-w) ,\qquad \theta_L(z) \theta_L(w) \sim + \log (z-w). 
\end{equation}
$\gamma$ can be understood as the coordinate on the boundary plane in the target space. We can also introduce the cylindrical coordinate
\begin{equation}
    u_L:=\log\gamma=-\varphi_L-\theta_L.
\end{equation}
This coordinate is null
\begin{equation}
    u_L(z)u_L(w) \sim 0.
\end{equation}
The translation on the cylinder corresponds to dilation and rotation on the plane. On the plane, the corresponding chiral current is
\begin{equation}
    J^3_{\pl}
    =
    -\frac1Q\p\Phi_L-\p\varphi_L
    =
    \p\chi_{\pl},
    \qquad
    \chi_{\pl}:=-\frac1Q\Phi_L-\varphi_L .
\end{equation}
Just like the relation between $L_0$ on the plane and the cylinder in CFT, the chiral current of translation on the cylinder is
\begin{equation}
    J^3_{\cy}
    =
    J^3_{\pl}
    -
    \frac{k}{4}\p u_L
    =
    \p\chi_{\cy},
\end{equation}
where
\begin{equation}
    \chi_{\cy}
    =
    -\frac1Q\Phi_L
    +
    \frac{k-4}{4}\varphi_L
    +
    \frac{k}{4}\theta_L .
\end{equation}
Then this boson is again null
\begin{equation}
    \chi_{\cy}(z)\chi_{\cy}(w)\sim0,
\end{equation}
because of
\begin{equation}
    \frac1{Q^2}
    +
    \z(\frac{k-4}{4}\y)^2
    -
    \z(\frac{k}{4}\y)^2
    =0.
\end{equation}
Moreover, $\chi_\cy$ generates translation along $u_L$
\begin{equation}
    \chi_{\cy}(z)u_L(w)\sim-\log(z-w).
\end{equation}
We complete \(\chi_{\cy}\) and \(u_L\) to a chiral basis by defining
\begin{equation}
    \sigma_L
    :=
    -\Phi_L
    +
    \frac1Q(\varphi_L+\theta_L).
\end{equation}
Then
\begin{equation}
    \sigma_L(z)\chi_{\cy}(w)\sim0,
    \qquad
    \sigma_L(z)u_L(w)\sim0, \qquad\sigma_L(z)\sigma_L(w)\sim-\log(z-w).
\end{equation}
The stress tensor in this basis is
\begin{equation}
    T
    =
    -:\p\chi_{\cy}\p u_L:
    -
    \frac12:(\p\sigma_L)^2:
    +
    \frac{Q}{2}\p^2\sigma_L .
\end{equation}
Therefore, the \((\chi_{\cy},u_L)\) system is a light-cone pair with no
background charge.  The background charge is carried only by \(\sigma_L\). We now introduce the right-moving analogues and define the real bosons
\begin{equation}
    X_L^+:=\chi_{\cy},
    \qquad
    X_L^-:=u_L,
    \qquad
    \Sigma_L:=\sigma_L,
\end{equation}
and
\begin{equation}
    X_R^+:=\bar\chi_{\cy},
    \qquad
    X_R^-:=\bar u_R,
    \qquad
    \Sigma_R:=\bar\sigma_R.
\end{equation}
The local real bosons are
\begin{equation}
    X^+:=X_L^+ + X_R^+,
    \qquad
    X^-:=X_L^- + X_R^-,
    \qquad
    \Sigma:=\Sigma_L+\Sigma_R.
\end{equation}
Thus, we have a decomposition of the worldsheet algebra
\begin{equation}
    O(3,3)
    \supset
    O(2,2)_{(X^+,X^-)}
    \oplus
    O(1,1)_\Sigma .
\end{equation}
The undeformed \(O(2,2)\) action is
\begin{equation}
    S^{(2,2)}_0
    =
    \frac1\pi\int d^2z~
    \frac12
    \z(
        \p X^+\bp X^-
        +
        \p X^-\bp X^+
    \y).
\end{equation}
The spectator is decoupled from this $O(2,2)$ system. The TsT deformation is the null \(O(2,2)\) deformation
\begin{equation}
    S_\mu^{(2,2)}
    =
    \frac1\pi\int d^2z~
    \left[
        \frac12
        \z(
            \p X^+\bp X^-
            +
            \p X^-\bp X^+
        \y)
        -
        \frac{2\mu}{k}\p X^+\bp X^+
    \right].
\end{equation}
Equivalently,
\begin{equation}
    E_\mu^{(2,2)}
    =
    \begin{pmatrix}
        -\frac{2\mu}{k} & \frac12\\
        \frac12 & 0
    \end{pmatrix}.
\end{equation}
This is the \(O(2,2)\) example in \eqref{action4.86}
\[
    \mu_{O(2,2)}\p\varphi_+\bp\varphi_+,
\]
with
\begin{equation}
    \varphi_+=X^+,
    \qquad
    \varphi_-=X^-,
    \qquad
    t=\frac12,
    \qquad
    \mu_{O(2,2)}=-\frac{2\mu}{k}.
\end{equation}
In the convention of this paper, we can choose for instance
\begin{equation}
    J_1^{(\mu)}
    =
    J^3_{\cy}
    =
    \p X_L^+, \qquad     \bar J_2^{(\mu)}
    =
    -\frac{2}{\pi k}\bar J^3_{\cy}
    =
    -\frac{2}{\pi k}\bp X_R^+ .
\end{equation}
Then
\begin{equation}
    \frac{\p S_\mu}{\p\mu}
    =
    \int d^2z~J_1^{(\mu)}\bar J_2^{(\mu)}
    =
    -\frac{2}{\pi k}
    \int d^2z~J^3_{\cy}\bar J^3_{\cy}.
\end{equation}
The TsT parameter \(\mu\) is already the flat coordinate $\mu = \lambda$. The local real boson dressing is
\begin{equation}
\begin{aligned}
    \Dr^{\ren}_{\mu}
    \z(X^{+,(\mu)}\y)
    &=
    X^{+,(0)} \\
    \Dr^{\ren}_{\mu}
    \z(X^{-,(\mu)}\y)
    &=
    X^{-,(0)}
    +
    \frac{2\mu}{k}X^{+,(0)} \\
    \Dr^{\ren}_{\mu}
    \z(\Sigma^{(\mu)}\y)
    &=
    \Sigma^{(0)}.
\end{aligned}
\end{equation}
The deformed stress tensor can be properly dressed back to the undeformed one:
\begin{equation}
\begin{aligned}
    T^{(\mu)} & = - \p X^{+,(\mu)}_L \p X^{-,(\mu)}_L + \frac{2\mu}{k} (\p X^{+,(\mu)}_L)^2 \\
    \Dr^\ren_\mu( T^{(\mu)} )&=  - \p X^{+,(0)}_L \p X^{-,(0)}_L.
\end{aligned}
\end{equation}
Dressing back to the undeformed theory gives
\begin{equation}\label{4.126}
\begin{aligned}
    \Dr^{\ren}_{\mu\to0}
    \z(u_L^{(\mu)}\y)
    &=
    u_L^{(0)}
    +
    \frac{2\mu}{k}\chi_{\cy}^{(0)} \\
        \Dr^{\ren}_{\mu\to0}
    \z(\bar u_R^{(\mu)}\y)
    &=
    \bar u_R^{(0)}
    +
    \frac{2\mu}{k}\bar\chi_{\cy}^{(0)}.
\end{aligned}
\end{equation}

Now, let us discuss the construction of the vertex operators. In the undeformed theory, the $w$-spectral-flow sector vertex operator in the Wakimoto representation is
\begin{equation}
    V^{(0)}_{j,m,\bar m;w} = e^{(\frac{w}{Q}-Qj)\Phi} e^{(w+m+j)\varphi_L + (m+j)\theta_L}e^{(w+\bar m+j)\varphi_R + (\bar m+j)\theta_R},
\end{equation}
or in terms of the $O(2,2)\oplus O(1,1)$ variables:
\begin{equation}
    V^{(0)}_{j,M,\bar M;w} = \exp \z( Qj \Sigma^{(0)} - w X^{+,(0)} - M_0 X^{-,(0)}_L - \bar M_0  X^{-,(0)}_R \y),
\end{equation}
where
\begin{equation}
    M_0 := m + \frac{k}{4}w,\qquad \bar M_0 := \bar m + \frac{k}{4}w.
\end{equation}
These are the charges of the cylinder currents
\begin{equation}
    J^{3,(0)}_{\cy}(z)V^{(0)}_{j,M,\bar M;w} (0)
    \sim
    \frac{M_0}{z}V^{(0)}_{j,M,\bar M;w},
    \qquad
    \bar J^{3,(0)}_{\cy}(\bar z)V^{(0)}_{j,M,\bar M;w}(0)
    \sim
    \frac{\bar M_0}{\bar z}V^{(0)}_{j,M,\bar M;w}(0).
\end{equation}
The worldsheet conformal weights of $V^{(0)}_{j,M,\bar M;w}$ are
\begin{equation}
    \begin{aligned}
        h_0 &= \frac{j(1-j)}{k-2} - M_0 w  \\
        \bar h_0 &= \frac{j(1-j)}{k-2} - \bar M_0 w.  \\
    \end{aligned}
\end{equation}
Now we study the deformed vertex. With our normalized deformation currents $J_1^{(\mu)}, \bar J_2^{(\mu)}$, the charges read
\begin{equation}
    Q_\mu=M_\mu,
    \qquad
    \bar Q_\mu=\frac{2}{\pi k}\bar M_\mu.
\end{equation}
According to \eqref{hq000000}, the flow of the charges and the conformal weights should be
\begin{equation}\label{4.133}
\begin{aligned}
    \p_\mu Q_\mu&=0,
    \qquad
    \p_\mu\bar Q_\mu=0 \\
    \p_\mu h_\mu
    &=
    -\pi Q_\mu\bar Q_\mu
    =
    -\frac{2}{k}M_\mu\bar M_\mu \\
        \p_\mu \bar h_\mu
    &=
    -\pi Q_\mu\bar Q_\mu
    =
    -\frac{2}{k}M_\mu\bar M_\mu.
\end{aligned}
\end{equation}
Therefore
\begin{equation}\label{4.134}
    h_\mu
    =
    h_0
    -
    \frac{2\mu}{k}M_0\bar M_0,
    \qquad
    \bar h_\mu
    =
    \bar h_0
    -
    \frac{2\mu}{k}M_0\bar M_0.
\end{equation}
In a spectral-flow sector with integer cylinder winding \(w\), one may write
\begin{equation}
\begin{aligned}
    h_\mu
    &=
    \frac{j(1-j)}{k-2}
    -
    M_0 w
    -
    \frac{2\mu}{k}M_0 \bar M_0 \\
     \bar h_\mu
    &=
    \frac{j(1-j)}{k-2}
    -
    \bar M_0 w
    -
    \frac{2\mu}{k}M_0 \bar M_0.
\end{aligned}
\end{equation}
However, as usual in string theory, the physical vertex should carry $h=\bar h=1$ in the old covariant quantization formalism. Thus, the parameters must be chosen so that the deformed vertex satisfies the physical-state conditions. In other words, in order to get an on-shell vertex $V^{(\mu)}_{j,M,\bar M;w}$ in the deformed theory, one has to start from an unphysical operator in the undeformed theory with some ``wrong" charges $M_\mu,\bar M_\mu$, so that the final conformal weights
\begin{equation}
\begin{aligned}
    h_\mu
    &=
    \frac{j(1-j)}{k-2}
    -
    M_\mu w
    -
    \frac{2\mu}{k}M_\mu \bar M_\mu \\
     \bar h_\mu
    &=
    \frac{j(1-j)}{k-2}
    -
    \bar M_\mu w
    -
    \frac{2\mu}{k}M_\mu \bar M_\mu
\end{aligned}
\end{equation}
are on-shell, namely
\begin{equation}
    M_0 w = M_\mu\z( w + \frac{2\mu}{k} \bar M_\mu  \y),\qquad \bar M_0 w = \bar  M_\mu\z( w + \frac{2\mu}{k} M_\mu  \y).
\end{equation}
This equation matches the deformed spectrum of the $T\bar T$-deformation of CFTs if we identify $M_\mu,\bar M_\mu$ as the deformed boundary energy and angular momentum \cite{Jiang:2019epa, Du:2024bqk}. Explicitly, the deformed exponential vertex with the charges $(M_0,\bar M_0)=(M_\mu,\bar M_\mu)$ and conformal weights in \eqref{4.134} is
\begin{equation}
    V^{(\mu)}_{j,M_\mu,\bar M_\mu,w} = e^{Qj\Sigma^{(\mu)} - M_\mu X^{-,(\mu)}_L - \bar M_\mu X^{-,(\mu)}_R - (w - \frac{2\mu}{k}(M_\mu -\bar M_\mu)  ) X^{+,(\mu)}_L - (w - \frac{2\mu}{k}(\bar M_\mu - M_\mu)  ) X^{+,(\mu)}_R}.
\end{equation}
The dressed operator is thus
\begin{equation}
    \Dr_\mu^\ren \z(  V^{(\mu)}_{j,M_\mu,\bar M_\mu,w}  \y) = e^{Qj\Sigma^{(0)} - w_L(\mu)X^{+,(0)}_L - w_R(\mu)X^{+,(0)}_R - M_\mu X^{-,(0)}_L - \bar M_\mu X^{-,(0)}_R },
\end{equation}
with
\begin{equation}
    w_L(\mu) = w + \frac{2\mu}{k} \bar M_\mu,\qquad w_R(\mu) = w + \frac{2\mu}{k}  M_\mu.
\end{equation}
This construction of vertices is consistent with \cite{Chakraborty:2019mdf}. The same construction can also be realized using the rule of non-local coordinates proposed in \cite{Azeyanagi:2012zd, Apolo:2019zai, Du:2024bqk, Cui:2023jrb}. Note that the dressed fields in \eqref{4.126} should be compared with the non-local coordinates defined in those works. The two constructions are conceptually different: the dressing map is based on invariance of correlation functions of an $O(2,2)$ realization, while the non-local coordinates are auxiliary fields for which the equation of motion of the undeformed theory is recovered. The non-local coordinates are fields in an auxiliary undeformed theory. In some sense, this auxiliary undeformed theory is the same as our $S_0$-theory, because the vertex operators in both theories are the same.

\section{Conclusion}

In this paper we have reformulated the conformal perturbation theory of
$J\bar J$-deformed CFTs in terms of a dressing of operators. The central
observation is that the Riemann bilinear identity converts the integrated
deformation operator
\begin{equation}
    \int_\Sigma d^2z~J_1(z)\bar J_2(\bar z)
\end{equation}
into a sum of cycle integrals. The local cycles around operator insertions
produce a dressing of the inserted operators, while the large cycles of a
higher-genus Riemann surface produce the deformation of the partition function.
This separates the local and global effects of the deformation in a way that is
well adapted to both conformal perturbation theory and modular properties.

At the level of local operators, the bare dressing obtained from the Riemann
bilinear identity reproduces the singular and regular terms of first-order
conformal perturbation theory. After applying the same renormalization
prescription as in conformal perturbation theory, it becomes a renormalized
linear map
\begin{equation}
    \Dr_\lambda^{\ren}: \mathcal H_\lambda \rightarrow \mathcal H_0 .
\end{equation}
This map transports operators from the deformed theory to the undeformed one.
In particular, it sends the deformed currents and stress tensor to the
undeformed currents and stress tensor. This gives an operator-level
interpretation of the fact that the deformation preserves the current algebra
and the central charge. The deformation is therefore encoded not in a new stress
tensor acting on the undeformed Hilbert space, but in a new set of dressed
operators inside the undeformed theory.

For partition functions with chemical potentials, the large-cycle part of the
Riemann bilinear identity gives closed flow equations. When the currents have
no self-interactions, the torus partition function obeys a heat equation,
\begin{equation}
    \p_\lambda Z_\lambda
    =
    4\pi^2\tau_2\p_\chi\p_{\bar\chi}Z_\lambda ,
\end{equation}
and the solution is a Gaussian convolution of the undeformed partition function.
We generalized this result to higher genus, where the kernel is controlled by
the imaginary part of the period matrix. When the currents have
self-interactions with levels $\kappa$ and $\bar\kappa$, the flow equation is
modified to
\begin{equation}
    \p_\lambda Z_\lambda
    =
    \left(
        4\pi^2\tau_2\p_\chi\p_{\bar\chi}
        -
        \pi\kappa\chi\p_{\bar\chi}
        -
        \pi\bar\kappa\bar\chi\p_\chi
    \right)Z_\lambda .
\end{equation}
We derived the analogous higher-genus equation and showed that, after an
appropriate non-holomorphic completion, the differential operator is modular
invariant. The integrated answer is a kernel of harmonic-oscillator type. Thus
the Riemann bilinear identity gives a uniform derivation of both the
deformation kernel and its modular covariance.

The same flow equation determines the deformation of individual states when
there is no mixing among states with the same quantum numbers, or when the
degenerate states have the same charge flow. The charges and conformal weights
obey
\begin{equation}
    \p_\lambda q=-\pi\kappa\bar q,\qquad
    \p_\lambda \bar q=-\pi\bar\kappa q,\qquad
    \p_\lambda h=\p_\lambda\bar h=-\pi q\bar q .
\end{equation}
These equations integrate to a hyperbolic rotation of the charges and a
corresponding shift of the conformal weights. In the no-self-interaction limit,
the charges are fixed and the conformal weights shift linearly by
$-\pi\lambda q\bar q$. Using the modular $S$-transformation, we argued that
dressed operators can be characterized as endpoint operators of defects. The
defect parameter is fixed by requiring that the resulting operator has the
deformed charges and conformal weights. This gives a practical criterion for
constructing dressed operators, especially when the currents can be bosonized.

The $O(2,2)$ examples provide explicit checks of the formalism. For two compact
bosons with a rank-one current-current deformation, the bare dressing of the
elementary real bosons preserves the OPE metric in the required way and maps
the normalized deformed currents back to the undeformed ones. Hence, in this
basis, the bare dressing coincides with the renormalized dressing up to the
field-strength factors that remove diagonal double contractions. The dressed
vertex operators have precisely the conformal weights and charges of the
corresponding operators in the deformed theory. These examples also show that
the flat coordinate of conformal perturbation theory can differ from the
coordinate appearing linearly in the sigma-model action, even in
no-self-interaction cases. Furthermore, descendant and non-exponential primary
operators require additional care: fixing the exponential part by charge flow is
not always enough, because the derivative structure must also be adjusted to
preserve primariness.

Finally, we applied the construction to strings on a TsT-deformed AdS$_3$
background. In the Wakimoto description, the relevant degrees of freedom can be
organized into an $O(2,2)\oplus O(1,1)$ system. The TsT deformation is then a
null current-current deformation in the $O(2,2)$ sector. The dressing of the
real bosons maps the deformed stress tensor back to the undeformed one and
produces deformed vertex operators whose conformal weights agree with the
expected TsT/single-trace deformation spectrum. This gives a direct
operator-level realization of the relation between worldsheet
current-current deformations and holographic $\TT$-type deformations.

There are several directions in which this framework can be developed further.
First, it would be useful to make the renormalization of the dressing map more
systematic beyond the examples considered here, especially for operators whose
OPEs with the currents contain higher-order poles. Second, the treatment of
degenerate spectra and operator mixing should be refined. The partition
function determines the flow of eigenvalues, but a full operator map in a
degenerate subspace requires additional data. Third, the defect-endpoint
criterion for dressed operators deserves a more intrinsic formulation, possibly
in terms of categories of line defects and their endpoints. Fourth, it would be
interesting to extend the analysis to non-abelian or non-semisimple current
algebras, where the deformation is generally not exactly marginal but may still
have controlled subsectors. Finally, the higher-genus kernels obtained here
suggest applications to string perturbation theory in TsT-deformed backgrounds
and to the study of correlation functions in single-trace $\TT$, $\JT$,
$T\bar J$, and $J\bar J$ holography.

\section*{Ackowledgements }
The author would like to thank Liangyu Chen, Zhengyuan Du, and Wei Song for useful discussions. K.L. is supported by the NSFC special fund for theoretical physics No. 12447108, No. 124B2094, and the National Key Research and Development Program of China No. 2020YFA0713000. 

\appendix

\bibliographystyle{JHEP}
\bibliography{ref.bib}

\providecommand{\href}[2]{#2}\begingroup\raggedright\begin{thebibliography}{10}

\bibitem{Borsato:2023dis}
R.~Borsato, \emph{{Lecture notes on current{\textendash}current deformations}}, \href{https://doi.org/10.1140/epjc/s10052-024-12966-5}{\emph{Eur. Phys. J. C} {\bfseries 84} (2024) 648} [\href{https://arxiv.org/abs/2312.13847}{{\ttfamily 2312.13847}}].

\bibitem{Chaudhuri:1988qb}
S.~Chaudhuri and J.~A. Schwartz, \emph{{A criterion for integrably marginal operators}}, \href{https://doi.org/10.1016/0370-2693(89)90393-6}{\emph{Phys. Lett. B} {\bfseries 219} (1989) 291}.

\bibitem{Forste:2003km}
S.~Forste and D.~Roggenkamp, \emph{{Current-current deformations of conformal field theories, and WZW models}}, \href{https://doi.org/10.1088/1126-6708/2003/05/071}{\emph{JHEP} {\bfseries 05} (2003) 071} [\href{https://arxiv.org/abs/hep-th/0304234}{{\ttfamily hep-th/0304234}}].

\bibitem{Fredenhagen:2007rx}
S.~Fredenhagen, M.~R. Gaberdiel and C.~A. Keller, \emph{{Symmetries of perturbed conformal field theories}}, \href{https://doi.org/10.1088/1751-8113/40/45/012}{\emph{J. Phys. A} {\bfseries 40} (2007) 13685} [\href{https://arxiv.org/abs/0707.2511}{{\ttfamily 0707.2511}}].

\bibitem{Borsato:2024cct}
R.~Borsato, \emph{{Lecture notes on current-current deformations}}, \href{https://doi.org/10.1140/epjc/s10052-024-12966-5}{\emph{Eur. Phys. J. C} {\bfseries 84} (2024) 648} [\href{https://arxiv.org/abs/2312.13847}{{\ttfamily 2312.13847}}].

\bibitem{Narain:1985jj}
K.~S. Narain, \emph{{New Heterotic String Theories in Uncompactified Dimensions < 10}}, \href{https://doi.org/10.1016/0370-2693(86)90682-9}{\emph{Phys. Lett. B} {\bfseries 169} (1986) 41}.

\bibitem{Narain:1986am}
K.~S. Narain, M.~H. Sarmadi and E.~Witten, \emph{{A Note on Toroidal Compactification of Heterotic String Theory}}, \href{https://doi.org/10.1016/0550-3213(87)90001-0}{\emph{Nucl. Phys. B} {\bfseries 279} (1987) 369}.

\bibitem{Giveon:1994fu}
A.~Giveon, M.~Porrati and E.~Rabinovici, \emph{{Target space duality in string theory}}, \href{https://doi.org/10.1016/0370-1573(94)90070-1}{\emph{Phys. Rept.} {\bfseries 244} (1994) 77} [\href{https://arxiv.org/abs/hep-th/9401139}{{\ttfamily hep-th/9401139}}].

\bibitem{Hassan:1992gi}
S.~F. Hassan and A.~Sen, \emph{{Marginal deformations of WZNW and coset models from O(d,d) transformation}}, \href{https://doi.org/10.1016/0550-3213(93)90429-S}{\emph{Nucl. Phys. B} {\bfseries 405} (1993) 143} [\href{https://arxiv.org/abs/hep-th/9210121}{{\ttfamily hep-th/9210121}}].

\bibitem{Lunin:2005jy}
O.~Lunin and J.~M. Maldacena, \emph{{Deforming field theories with U(1) x U(1) global symmetry and their gravity duals}}, \href{https://doi.org/10.1088/1126-6708/2005/05/033}{\emph{JHEP} {\bfseries 05} (2005) 033} [\href{https://arxiv.org/abs/hep-th/0502086}{{\ttfamily hep-th/0502086}}].

\bibitem{Sfondrini:2019smd}
A.~Sfondrini and S.~J. van Tongeren, \emph{{$T\bar T$ deformations as TsT transformations}}, \href{https://doi.org/10.1103/PhysRevD.101.066022}{\emph{Phys. Rev. D} {\bfseries 101} (2020) 066022} [\href{https://arxiv.org/abs/1908.09299}{{\ttfamily 1908.09299}}].

\bibitem{Zamolodchikov:2004ce}
A.~B. Zamolodchikov, \emph{{Expectation value of composite field $T\bar T$ in two-dimensional quantum field theory}},  \href{https://arxiv.org/abs/hep-th/0401146}{{\ttfamily hep-th/0401146}}.

\bibitem{Smirnov:2016lqw}
F.~A. Smirnov and A.~B. Zamolodchikov, \emph{{On space of integrable quantum field theories}}, \href{https://doi.org/10.1016/j.nuclphysb.2016.12.014}{\emph{Nucl. Phys. B} {\bfseries 915} (2017) 363} [\href{https://arxiv.org/abs/1608.05499}{{\ttfamily 1608.05499}}].

\bibitem{Cavaglia:2016oda}
A.~Cavaglia, S.~Negro, I.~M. Szecsenyi and R.~Tateo, \emph{{$T\bar T$-deformed 2D Quantum Field Theories}}, \href{https://doi.org/10.1007/JHEP10(2016)112}{\emph{JHEP} {\bfseries 10} (2016) 112} [\href{https://arxiv.org/abs/1608.05534}{{\ttfamily 1608.05534}}].

\bibitem{McGough:2016lol}
L.~McGough, M.~Mezei and H.~Verlinde, \emph{{Moving the CFT into the bulk with $T\bar T$}}, \href{https://doi.org/10.1007/JHEP04(2018)010}{\emph{JHEP} {\bfseries 04} (2018) 010} [\href{https://arxiv.org/abs/1611.03470}{{\ttfamily 1611.03470}}].

\bibitem{Giveon:2017nie}
A.~Giveon, N.~Itzhaki and D.~Kutasov, \emph{{$ \mathrm{T}\overline{\mathrm{T}} $ and LST}}, \href{https://doi.org/10.1007/JHEP07(2017)122}{\emph{JHEP} {\bfseries 07} (2017) 122} [\href{https://arxiv.org/abs/1701.05576}{{\ttfamily 1701.05576}}].

\bibitem{Giveon:2017myj}
A.~Giveon, N.~Itzhaki and D.~Kutasov, \emph{{A solvable irrelevant deformation of $AdS_3/CFT_2$}}, \href{https://doi.org/10.1007/JHEP12(2017)155}{\emph{JHEP} {\bfseries 12} (2017) 155} [\href{https://arxiv.org/abs/1707.05800}{{\ttfamily 1707.05800}}].

\bibitem{Chakraborty:2019mdf}
S.~Chakraborty, A.~Giveon and D.~Kutasov, \emph{{$T\bar{T}$, $J\bar{T}$, $T\bar{J}$ and String Theory}}, \href{https://doi.org/10.1088/1751-8121/ab3710}{\emph{J. Phys. A} {\bfseries 52} (2019) 384003} [\href{https://arxiv.org/abs/1905.00051}{{\ttfamily 1905.00051}}].

\bibitem{Apolo:2019zai}
L.~Apolo, S.~Detournay and W.~Song, \emph{{TsT, $T\bar T$ and black strings}}, \href{https://doi.org/10.1007/JHEP06(2020)109}{\emph{JHEP} {\bfseries 06} (2020) 109} [\href{https://arxiv.org/abs/1911.12359}{{\ttfamily 1911.12359}}].

\bibitem{Hashimoto:2019wct}
A.~Hashimoto and D.~Kutasov, \emph{{$ T\overline{T},J\overline{T},T\overline{J} $ partition sums from string theory}}, \href{https://doi.org/10.1007/JHEP02(2020)080}{\emph{JHEP} {\bfseries 02} (2020) 080} [\href{https://arxiv.org/abs/1907.07221}{{\ttfamily 1907.07221}}].

\bibitem{Apolo:2021wcn}
L.~Apolo and W.~Song, \emph{{TsT, black holes, and $T\bar T + J\bar T + T\bar J$}}, \href{https://doi.org/10.1007/JHEP04(2022)177}{\emph{JHEP} {\bfseries 04} (2022) 177} [\href{https://arxiv.org/abs/2111.02243}{{\ttfamily 2111.02243}}].

\bibitem{Guica:2017lia}
M.~Guica, \emph{{An integrable Lorentz-breaking deformation of two-dimensional CFTs}}, \href{https://doi.org/10.21468/SciPostPhys.5.5.048}{\emph{SciPost Phys.} {\bfseries 5} (2018) 048} [\href{https://arxiv.org/abs/1710.08415}{{\ttfamily 1710.08415}}].

\bibitem{Guica:2021fkv}
M.~Guica, \emph{{A definition of primary operators in $J\bar T$-deformed CFTs}}, \href{https://doi.org/10.21468/SciPostPhys.13.3.045}{\emph{SciPost Phys.} {\bfseries 13} (2022) 045} [\href{https://arxiv.org/abs/2112.14736}{{\ttfamily 2112.14736}}].

\bibitem{Cui:2023jrb}
W.~Cui, H.~Shu, W.~Song and J.~Wang, \emph{{Correlation functions in the TsT/$T\bar T$ correspondence}}, \href{https://doi.org/10.1007/JHEP04(2024)017}{\emph{JHEP} {\bfseries 04} (2024) 017} [\href{https://arxiv.org/abs/2304.04684}{{\ttfamily 2304.04684}}].

\bibitem{Du:2024bqk}
Z.~Du, W.-X. Lai, K.~Liu and W.~Song, \emph{{Asymptotic symmetries in the $TsT/T\bar{T}$ correspondence}}, \href{https://doi.org/10.21468/SciPostPhys.18.2.049}{\emph{SciPost Phys.} {\bfseries 18} (2025) 049} [\href{https://arxiv.org/abs/2407.19495}{{\ttfamily 2407.19495}}].

\bibitem{Moriwaki:2020zyl}
Y.~Moriwaki, \emph{{Two-dimensional conformal field theory, full vertex algebra and current-current deformation}}, \href{https://doi.org/10.1016/j.aim.2023.109125}{\emph{Adv. Math.} {\bfseries 427} (2023) 109125} [\href{https://arxiv.org/abs/2007.07327}{{\ttfamily 2007.07327}}].

\bibitem{Giribet:2020mkz}
G.~Giribet and M.~Leoni, \emph{{Current-current deformations, conformal integrals and correlation functions}}, \href{https://doi.org/10.1007/JHEP04(2020)194}{\emph{JHEP} {\bfseries 04} (2020) 194} [\href{https://arxiv.org/abs/2003.02864}{{\ttfamily 2003.02864}}].

\bibitem{Cardy:2019qao}
J.~Cardy, \emph{{$T\bar T$ deformation of correlation functions}}, \href{https://doi.org/10.1007/JHEP12(2019)160}{\emph{JHEP} {\bfseries 12} (2019) 160} [\href{https://arxiv.org/abs/1907.03394}{{\ttfamily 1907.03394}}].

\bibitem{Kruthoff:2020hsi}
J.~Kruthoff and O.~Parrikar, \emph{{On the flow of states under $T\overline{T}$}},  \href{https://arxiv.org/abs/2006.03054}{{\ttfamily 2006.03054}}.

\bibitem{Giveon:1998ns}
A.~Giveon, D.~Kutasov and N.~Seiberg, \emph{{Comments on string theory on $AdS_3$}}, \href{https://doi.org/10.4310/ATMP.1998.v2.n4.a3}{\emph{Adv. Theor. Math. Phys.} {\bfseries 2} (1998) 733} [\href{https://arxiv.org/abs/hep-th/9806194}{{\ttfamily hep-th/9806194}}].

\bibitem{Maldacena:2000hw}
J.~Maldacena and H.~Ooguri, \emph{{Strings in $AdS_3$ and the $SL(2,R)$ WZW model. Part 1: The spectrum}}, \href{https://doi.org/10.1063/1.1377273}{\emph{J. Math. Phys.} {\bfseries 42} (2001) 2929} [\href{https://arxiv.org/abs/hep-th/0001053}{{\ttfamily hep-th/0001053}}].

\bibitem{Maldacena:2000kv}
J.~Maldacena, H.~Ooguri and J.~Son, \emph{{Strings in $AdS_3$ and the $SL(2,R)$ WZW model. Part 2: Euclidean black hole}}, \href{https://doi.org/10.1063/1.1377039}{\emph{J. Math. Phys.} {\bfseries 42} (2001) 2961} [\href{https://arxiv.org/abs/hep-th/0005183}{{\ttfamily hep-th/0005183}}].

\bibitem{Maldacena:2001km}
J.~Maldacena and H.~Ooguri, \emph{{Strings in $AdS_3$ and the $SL(2,R)$ WZW model. Part 3: Correlation functions}}, \href{https://doi.org/10.1103/PhysRevD.65.106006}{\emph{Phys. Rev. D} {\bfseries 65} (2002) 106006} [\href{https://arxiv.org/abs/hep-th/0111180}{{\ttfamily hep-th/0111180}}].

\bibitem{Knighton:2023mhq}
B.~Knighton, S.~Seet and V.~Sriprachyakul, \emph{{Spectral flow and localisation in AdS$_{3}$ string theory}}, \href{https://doi.org/10.1007/JHEP05(2024)113}{\emph{JHEP} {\bfseries 05} (2024) 113} [\href{https://arxiv.org/abs/2312.08429}{{\ttfamily 2312.08429}}].

\bibitem{cardy1996scaling}
J.~Cardy, \emph{Scaling and renormalization in statistical physics}, vol.~5. Cambridge university press, 1996.

\bibitem{Callebaut:2019omt}
N.~Callebaut, J.~Kruthoff and H.~Verlinde, \emph{{$ T\overline{T} $ deformed CFT as a non-critical string}}, \href{https://doi.org/10.1007/JHEP04(2020)084}{\emph{JHEP} {\bfseries 04} (2020) 084} [\href{https://arxiv.org/abs/1910.13578}{{\ttfamily 1910.13578}}].

\bibitem{griffiths2014principles}
P.~Griffiths and J.~Harris, \emph{Principles of algebraic geometry}. John Wiley \& Sons, 2014.

\bibitem{Eynard:2018tcr}
B.~Eynard, \emph{{Lectures notes on compact Riemann surfaces}},  \href{https://arxiv.org/abs/1805.06405}{{\ttfamily 1805.06405}}.

\bibitem{Zamolodchikov:1986gt}
A.~B. Zamolodchikov, \emph{{Irreversibility of the Flux of the Renormalization Group in a 2D Field Theory}}, {\emph{JETP Lett.} {\bfseries 43} (1986) 730}.

\bibitem{Sen:2017gfr}
K.~Sen and Y.~Tachikawa, \emph{{First-order conformal perturbation theory by marginal operators}},  \href{https://arxiv.org/abs/1711.05947}{{\ttfamily 1711.05947}}.

\bibitem{Chakraborty:2024mls}
S.~Chakraborty, A.~Giveon and A.~Hashimoto, \emph{{Thermal partition function of $ {J}_3{\overline{J}}_3 $ deformed AdS$_{3}$}}, \href{https://doi.org/10.1007/JHEP07(2024)141}{\emph{JHEP} {\bfseries 07} (2024) 141} [\href{https://arxiv.org/abs/2403.03979}{{\ttfamily 2403.03979}}].

\bibitem{Eguchi:1986sb}
T.~Eguchi and H.~Ooguri, \emph{{Conformal and Current Algebras on General Riemann Surface}}, \href{https://doi.org/10.1016/0550-3213(87)90686-9}{\emph{Nucl. Phys. B} {\bfseries 282} (1987) 308}.

\bibitem{Jiang:2019epa}
Y.~Jiang, \emph{{A pedagogical review on solvable irrelevant deformations of 2D quantum field theory}}, \href{https://doi.org/10.1088/1572-9494/abe4c9}{\emph{Commun. Theor. Phys.} {\bfseries 73} (2021) 057201} [\href{https://arxiv.org/abs/1904.13376}{{\ttfamily 1904.13376}}].

\bibitem{Azeyanagi:2012zd}
T.~Azeyanagi, D.~M. Hofman, W.~Song and A.~Strominger, \emph{{The Spectrum of Strings on Warped AdS$_3 \times$ S$^3$}}, \href{https://doi.org/10.1007/JHEP04(2013)078}{\emph{JHEP} {\bfseries 04} (2013) 078} [\href{https://arxiv.org/abs/1207.5050}{{\ttfamily 1207.5050}}].

\end{thebibliography}\endgroup

\end{document}